\documentclass[twocolumn]{aastex62}

\accepted{26th May, 2023 ApJ}

\shorttitle{}
\shortauthors{}

\usepackage{graphicx}
\usepackage{epsfig}
\usepackage{epstopdf}
\usepackage{times}
\usepackage[]{natbib}
\usepackage{url}
\usepackage{color}
\usepackage{amssymb,amsmath}
\usepackage{xfrac}
\usepackage{float}

\usepackage[varg]{txfonts}
\usepackage{color}           
\usepackage{lineno}

\usepackage{enumitem}
\setlist[enumerate]{itemsep=0mm}

\newcommand{\pder}[2][]{\frac{\partial#1}{\partial#2}}
\newcommand{\lder}[2][]{\frac{D #1}{D #2}}

\begin{document}

\title{ Effects of partial ionization on magnetic flux emergence in the Sun.}

\correspondingauthor{Georgios Chouliaras}
\email{gc205@st-andrews.ac.uk}

\author{Georgios Chouliaras}
\affil{School of Mathematics and Statistics, St. Andrews University, St. Andrews, KY16 9SS, UK}

\author{P.Syntelis}
\affil{Department of Physics, University of Ioannina, 45110, Ioannina, Greece}

I \author{V.Archontis}
\affil{Department of Physics, University of Ioannina, 45110, Ioannina, Greece}
\affil{School of Mathematics and Statistics, St. Andrews University, St. Andrews, KY16 9SS, UK}

 \begin{abstract}
 We have performed 3-D numerical simulations to investigate the effect of partial ionization on the process of magnetic flux emergence. In our study, we have modified the single-fluid MHD equations to include the presence of neutrals and have performed two basic experiments: one that assumes a fully ionized plasma (FI case) and one that assumes a partially ionized plasma (PI case). We find that the PI case brings less dense plasma to and above the solar surface. Furthermore, we find that partial ionization alters the emerging magnetic field structure, leading to a different shape of the polarities in the emerged bipolar regions compared to the FI case. The amount of emerging flux into the solar atmosphere is larger in the PI case, which has the same initial plasma beta as the FI case, but a larger initial magnetic field strength. The expansion of the field above the photosphere occurs relatively earlier in the PI case, and we confirm that the inclusion of partial ionization reduces cooling due to adiabatic expansion. However, it does not appear to work as a heating mechanism for the atmospheric plasma. The performance of these experiments in three dimensions shows that PI does not prevent the formation of unstable magnetic structures, which erupt into the outer solar atmosphere.

 \end{abstract}

 \keywords{Sun: activity, Sun: interior,
                  Sun: Magnetic fields, Magnetohydrodynamics,  partial ionization (MHD), methods: numerical
               }

\section{Introduction} \label{sec:intro}
\par Magnetic flux emergence is a fundamental process in the Sun, which plays a key role on the driving and/or triggering of many dynamical phenomena, such as the formation of active regions, the onset of flares, jets, eruptions, etc.  In the past, magnetic flux emergence has been studied in detail, via 3D MHD numerical simulations \citep[e.g.,][]{Moreno-Insertis_etal1996,Magara_etal2001,Fan_2001,Manchester_etal2004,Archontis_etal2004} and references therein). The vast majority of these experiments have used a solar atmosphere assuming a fully ionized plasma. 
However, it is well known, that the solar photosphere and chromosphere are partially ionized, which brings the presence of electrons and neutrals in addition to the existence of ions in the atmospheric plasma. 
\par \citet{Leake_etal2006} performed 2.5D MHD simulations  of magnetic flux emergence including the presence of neutrals in the solar atmosphere. More precisely, they have used the generalised Ohms law for a fluid containing three different species given by \citet{1965RvPP....1..205B} and they have neglected pressure and Hall term. 
They found that the rate of flux emergence in the solar atmosphere increased due to the effect of partial ionization. In addition, the emerging magnetic field appeared to 
be more diffusive. Another important result of this study was that, due to the perpendicular resistivity induced by the ion-neutral collisions, 
the currents that emerge to the corona were aligned with the field creating a force-free coronal field. We have to note that this model used a simple equation 
of state without the implementation of the ionization-recombination effect. \citet{Arber_etal2007} extended the afore-mentioned model from 2.5D to 3D and they confirmed 
the results from the 2.5D simulations, highlighting that perpendicular resistivity successfully dissipates perpendicular current. Furthermore, this study showed 
that when partial ionization is considered, then less chromospheric material is lifted to the corona. \citet{Leake_etal2013b} modified their previous 2.5D model 
by taking into account the ionization-recombination effect in the equation of state. They performed simulations of the emergence of magnetic flux tubes, with 
different radii and initial axial flux. They found that the addition of ionization-recombination affects the rise speed of the tube's axis, which reaches smaller heights 
at the end of the simulations. Furthermore, this study showed that the energy in the sheared component of the magnetic field is reduced compared to the 
fully ionized case. In fact, it was reported that the tubes in the PI experiments lift up to 89 \% less material than the fully ionized case. 
An interesting result of this study was that the implementation of the partial ionization to the equation of state led to the reduction of the 
free magnetic energy supplied to the corona. It was reported that the latter does not favour the formation of coronal structures that might become unstable and erupt.
\par In our study, we have extended the \citet{Leake_etal2013b} model from 2.5-D to 3-D. Firstly, we confirm some of the results from their work. 
Secondly, we report new results as a consequence of the extension of this model to 3-D. Finally, we briefly discuss the recurrent eruptions, which occur 
after the emergence at the solar atmosphere. The structure of our paper is as follows: We present the model in chapter 2, then  in chapter 3 we present 
the results and finally in chapter 4 we report our conclusions.

\section{Model}

\subsection{Equations}
We numerically solve the 3D time-dependent, compressible, resistive MHD equations in Cartesian geometry using the Lare3D code of \citet{Arber_etal2001}. We modify the 
equations to include the effects of partial ionization (neglecting Hall term). For that, we follow the 2.5D model of \citet{Leake_etal2013b}.

We assume populations of ions ($i$), neutrals ($n$) and electrons ($e$). Therefore, the total mass density ($\rho$), gas pressure ($P$) and 
specific energy ($\epsilon$) are obtained by summing over the three species. For instance, $\rho = \sum_k m_k n_k$, where $k=i,e,n$ and $m_k$ and $n_k$ are the 
mass and number density of the species $k$. 

The resulting MHD equations solved are:
\begin{align}
\lder[\rho]{t} &= -\rho \nabla\cdot\mathbf{v}, \label{eq:cont}\\
\rho \lder[\mathbf{v}]{t}  &=  - \nabla P 
                        + \mathbf{j} \times \mathbf{B} 
                        - g_0 \mathbf{\hat{z}}
                        + \mathbf{S}_{visc} , \label{eq:momentum}\\
\lder[\mathbf{B}]{t} &= -\mathbf{B}\left( \nabla \cdot \mathbf{v} \right)
                        +\left( \mathbf{B} \cdot \nabla \right) \mathbf{v} 
                        - \nabla \times\left[\eta \mathbf{j_\parallel} + (\eta+\eta_\perp) \mathbf{j_\perp}\right], \label{eq:induction}\\
\rho\lder[\epsilon]{t} &= -P \nabla\cdot\mathbf{v} 
                        + \eta j^2_\parallel + (\eta+\eta_\perp) j^2_\perp
                        + Q_{visc},\label{eq:energy}
\end{align}
where $\mathbf{v}$, $\mathbf{B}$, $\rho$, $P$ are namely the  velocity vector, magnetic field vector, density and gas pressure.
Gravity is $g_0=274$~m s$^{-1}$. 
Viscosity is added through $\mathbf{S}_{visc}= \pder[\sigma_{ij}]{x_j} \mathbf{\hat{e}}_i$ and $Q_{visc} = \varepsilon_{ij}\sigma_{ij}$,
 where
$\sigma_{ij} = 2 \nu \left( \varepsilon_{ij} - \frac{1}{3}\delta_{ij} \nabla \cdot \mathbf{v} \right)$ and $\varepsilon_{ij} = \frac{1}{2} \left( \pder[v_i]{x_j} + \pder[v_j]{x_i} \right)$, and $\nu=622$~kg m$^{-1}$ s$^{-1}$ (10~$^{-2}$ in non-dimensional units).
The current density is treated using its components parallel ($\mathbf{j}_\parallel$) and perpendicular ($\mathbf{j}_\perp$) to the magnetic field vector. These are defined as
\begin{align}
    \mathbf{j}_\parallel = \frac{ ( \mathbf{j} \cdot \mathbf{B }) }{B^2} 
    \quad \textnormal{and} \quad
    \mathbf{j}_\perp = \frac{  \mathbf{B} \times (\mathbf{j} \times \mathbf{B }) }{B^2},
\end{align} 
where $\mathbf{j}= \frac{1}{\mu_0} \nabla \times \mathbf{B}$ is the full current density vector. 

We do not calculate the uniform resistivity from the collision frequencies of electrons with ions and neutrals as \citet{Leake_etal2013b}. Instead, we use a constant uniform resistivity of $\eta=4.6$~$\Omega$~m (10$^{-2}$ in non-dimensional units). That way, we focus only on the effects of the perpendicular resistivity term ($\eta_\perp$). The perpendicular resistivity due to the ambipolar diffusion of neutrals is given by 
\begin{align}
  \eta_\perp = \frac{\xi_n B^2}{\alpha_n},
\end{align}
where $\alpha_n = m_e n_e \nu^\prime_{en} + m_i n_i \nu^\prime_{in}$ is calculated using the effective collisional frequencies for electron-neutral collisions ($\nu^\prime_{en}$) and for ion-neutral collisions ($\nu^\prime_{in}$). These terms are calculated according to \citet{Leake_etal2005}. 
The neutral fraction ($\xi_n$) is defined as
\begin{align}
    \xi_n = \frac{n_n}{n_n+n_i}.
\end{align}
Just for clarity, we note here that the expressions and names used for the perpendicular resistivity vary in the literature. For instance, \citet{Leake_etal2013b} refer to it as Pedersen resistivity ($\eta_p=\eta+\eta_\perp$), grouping the resistivity terms of the current density perpendicular to the field (Eq.~\ref{eq:induction},~\ref{eq:energy}) in one term. \citet{Arber_etal2007} refer to the perpendicular resistivity as Cowling resistivity ($\eta_c = \eta_\perp$), while \cite{Martinez-Sykora_etal2012} refer to it as ambipolar diffusion resistivity ($\eta_{amb}=n_\perp$). 
Irrespective of the name used or the grouping of the terms, eventually, the terms used to describe the resistive effects in the induction and energy equation are the same. Effectively, there is little difference between Pedersen, Cowling and ambipolar diffusion resistivity.

We also solve the ideal gas law, which is
\begin{align}
    P &= \frac{\rho k_B T}{\mu_m},\quad \textnormal{where} \quad \mu_m=\frac{m_i}{2-\xi_n}.
    \label{eq:ideal_gas_law}
\end{align}
For the reduced mass $\mu_m$, we use $m_i= m_f m_p$, where $m_p$ is the mass of proton and $m_f=1.25$, $k_B$ is Boltzmann constant. 

For the specific internal energy equation, we take into account changes both to the temperature and to the ionization level,
\begin{align}
    \epsilon &= \frac{k_B T}{\mu_m (\gamma -1)} + (1-\xi_n) \frac{X_i}{m_i},
    \label{eq:specific_energy}
\end{align}
where  $\gamma=5/3$ and $X_i=13.6$~eV is the first ionization energy of hydrogen. When performing simulations without the effects of partial ionization ($\xi_n=0$), we omit the second term of the equation and keep only the gas term to compare the results with previous simulations of fully ionized plasma.

We calculate $\xi_n$ according to \citet{Leake_etal2013b}. Briefly, we use the Saha equation \cite{Saha_1921} below the photosphere (which is in local thermodynamic equilibrium) and the modified Saha equation \cite{Brown_1973} above the photosphere (which is not in local thermodynamic equilibrium), to derive the $n_i^2/n_n$ ratio. Afterwards, the neutral fraction $\xi_n$ is calculated.

The above set of equations is normalized using the photospheric values of density $\rho_\mathrm{u}=1.67 \times 10^{-7}$ g cm$^{-3}$, length $H_\mathrm{u}=180$~km and magnetic field strength $B_\mathrm{c}=300$~G. Using the value of $m_i$ mentioned before, we derive
from these we get temperature $T_\mathrm{u}=649$~K, velocity $v_\mathrm{u}=2.1\ \mathrm{km} \ \mathrm{s}^{-1}$ and time $t_\mathrm{u}=86.9\ \mathrm{s}$.

\subsection{Initial conditions}

We use a numerical domain the same as the one used in \citet{Syntelis_etal2017}.
The computational domain has a physical size of $64.8^3 \mathrm{Mm}$ in a $420^3$ uniform grid. 
The interior extends at $-7.2$~Mm$\le z < 0$~Mm, the  photospheric-chromospheric layer at $0$~Mm$\le z < 1.8$~Mm, the transition region at $1.8$~Mm$\le z < 3.2$~Mm and an isothermal corona at $3.2$~Mm$\le z < 57.6$~Mm.
We assume periodic boundary conditions in the $y$ direction. Open boundary conditions are used in the $x$ direction and at the top of the numerical box. The bottom boundary is set to be closed. We place a cylindrical twisted flux tube (FT) $2.3 Mm$ below photosphere.

We use two simulations in this study as mentioned above (FI and PI).

The background of both simulations is comprised by an adiabatically stratified solar interior (making the interior marginally stable to convection) 
and a stratified model atmosphere, both in hydrostatic equilibrium. We note that the reduced mass for the FI simulation is $\mu_m=m_i$ instead of $\mu_m=m_i/2$. That 
way we take into account the mass of the neutrals and get realistic coronal densities.

For both simulations, we assume the temperature of the atmosphere ($z>0$) to follow a tangential temperature profile, 
\begin{align}
    T(z) = T_{ph} + \frac{T_{cor}- T_{ph}}{2} 
            \left( \tanh{\frac{z-z_{cor}}{w_{tr}} +1} \right),
\end{align}
where $T_{ph}=6360$~K, $T_{cor} = 0.95$~MK, $z_{cor}=2.52$~Mm and $w_{tr}=0.18$~Mm. Having set this atmospheric temperature profile, we numerically 
solve the hydrostatic equation $dP/dz = - gz$ to derive the density of the atmosphere. We get a photospheric density of $\rho_{ph}= 1.67\times10^{-4}$~kg m$^{-3}$. For PI, we also calculate the neutral fraction $\xi_n$ of the resulting $T$ and $\rho$ profiles.

To setup the initial stratification below the photosphere ($z<0$), we require the vertical temperature gradient, assuming that a 
plasma element moves adiabatically. That is estimated using energy Eq.~\ref{eq:specific_energy}. For the FI simulation ($\xi_n=0$) we omit the 
second term of Eq.~\ref{eq:specific_energy}, which simplifies the equation significantly. From that equation, the temperature profile for FI is:
\begin{align}
    \left( \frac{dT}{dz}\right)_a = - \frac{\mu_m g}{k_B} \frac{\gamma-1}{\gamma}.
\end{align}
This equation is solved analytically and then the density is calculated by solving numerically the hydrostatic equation. The boundary conditions are $T_{ph}$ and $\rho_{ph}$. This stratification is used in all FI flux emergence simulations \citep[e.g.][]{Fan_2001, Manchester_etal2004, Archontis_etal2004, 
Moreno-Insertis_etal2008, Toriumi_etal2011, Leake_etal2013a,Syntelis_etal2015}.

However, in PI, the full Eq.~\ref{eq:specific_energy} is needed in order to derive the temperature profile of the solar interior. Otherwise, the 
PI interior will not be adiabatically and convectively stable. Taking into account the full Eq.~\ref{eq:specific_energy}, \citet{Leake_etal2013b} showed that 
the temperature profile for the solar interior becomes:
\begin{align}
    \left( \frac{dT}{dz}\right)_a = - \frac{\mu_m g}{k_B} \frac{\gamma-1}{\gamma}
            \left(
                \frac{ 1+\theta \frac{\zeta(1-\zeta)}{2} }
                {\frac{\gamma}{\gamma-1} + \theta^2 \frac{\zeta(1-\zeta)}{2} } 
            \right),
\label{eq:stratification}            
\end{align}
where 
\begin{align}
\theta = \frac{\gamma}{\gamma-1} + \frac{X_i}{k_B T}, 
    \quad \textnormal{and} \quad
\zeta=1-\xi_n.
\end{align}
The method to solve this equation is described in \citet{Leake_etal2013b}. Briefly, we initially set $\zeta=1$ (i.e. assume the plasma to be FI) and use as boundary conditions $T_{ph}$ and $\rho_{ph}$. We use a 4th order Runge-Kutta scheme to solve Eq.~\ref{eq:stratification} for $z<0$ and then find the 
density by numerically solving the hydrostatic equation.  Using the Saha equation, we calculate for the current $T$ and $\rho$ a new ionization fraction, $\xi_n$, and then the new $\zeta$ and $\theta$. Then, we iterate these steps, until $\zeta$ converges for a set of $T$, $\rho$ in the interior.

\begin{figure}
\centering
\includegraphics[width=\columnwidth]{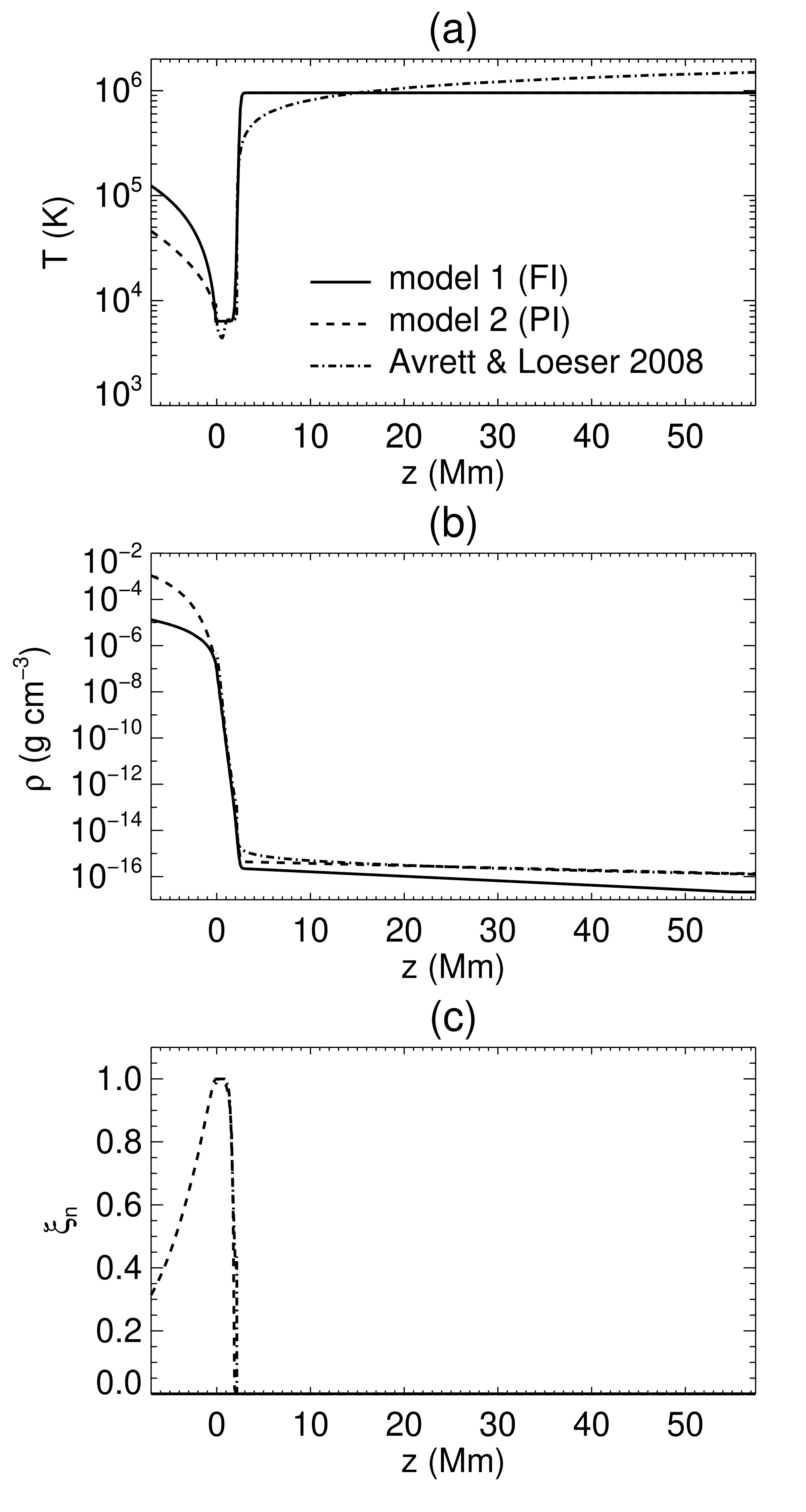}
\caption{
The (a) temperature stratification, (b) density stratification and (c) the neutral fraction of the solar interior and the atmosphere of the simulations. The solid line is FI (solid) and the dashed line is PI (dashed). Over plotted with a dashed-dot line is model C7 of \citet{Avrett_loeser2008} for comparison. 
}
\label{fig:stratification}
\end{figure}

The resulting temperature, density and neutral fraction of the stratified interior and atmosphere is shown in Fig.~\ref{fig:stratification}. We overplot 
the same parameters from the model atmosphere C7 of \citet{Avrett_loeser2008} for comparison.

Inside the interior, we place a horizontal cylindrical twisted flux tube along the $y$-axis. 
The magnetic field of the flux tube is defined as
\begin{align}
B_{y} &=B_\mathrm{0} \exp(-r^2/R^2), \\
B_{\phi} &= \alpha r B_{y},
\end{align}
where $R=450$~km is a measure of the FT's radius, $r$ is the radial distance from the FT's axis and $\alpha/2\pi$ is the twist per unit of length. We set $\alpha= 0.4$ ($0.0023$~km$^{-1}$), which makes the flux tube highly twisted but kink stable.

For FI, we set the magnetic field's strength to $B_0=3150$~G ($\beta=18.4$), same to \citet{Syntelis_etal2017}. That way, the FI simulation is similar to the one described in detail in \citet{Syntelis_etal2017} and direct comparisons can be made.
Notice in Fig.~\ref{fig:stratification} that the FI and PI background stratifications have different densities and temperatures in the interior, as it results from solving two different equations. For PI, we, therefore, do not use the same $B_0$ as FI, but the same $\beta$. That results to $B_0=7882$~G for PI.

To make the flux tube emerge, we change the density profile inside the flux tube. For FI, we use the usual method \cite[e.g.][]{Fan_2001}. 
The background solar interior has a pressure, temperature and density profile of $P_0$, $T_0$ and $\rho_0$. By adding the flux tube, we add an additional pressure excess due to the magnetic field. Firstly, we require this pressure excess to be in radial force equilibrium $(dP_{exc}/dr) = (\mathbf{j}\times\mathbf{B})\mathbf{\hat{e}}_r$, finding the equation of $P_{exc}$ to be \citep[see ][for details]{Murray_etal2006}:
\begin{align}
    P_{exc} = \frac{1}{2\mu}\left[ \alpha^2 \left( \frac{R^2}{2} -r^2 \right) -1 \right] B_y^2
\end{align}
Then, we require the flux tube to be in pressure equilibrium. The gas pressure in the interior of the tube, $P_i$, becomes $P_i=P_0 - P_{exc}$. Because inside the solar interior the isotropic thermal conduction is large, we expect the magnetized plasma to be in thermal equilibrium with the background non-magnetized plasma. We therefore set the flux tube to be in thermal equilibrium with the background ($T_i=T_0$). That leads to a density deficit ($\Delta\rho=\rho_i-\rho_0$) in the interior of the flux tube of $\Delta\rho = - \rho_0 P_{exc}/P_0$ that makes the flux tube buoyant. To avoid emerging the whole length of the flux tube, we reduce the density deficit towards the flanks of the flux tube by
\begin{align}
    \Delta\rho = - \rho_0 \frac{P_{exc}}{P_0} e^{-y^2/\lambda^2},
    \label{eq:density_deficit}
\end{align}
where $\lambda$ is thus a measure of the length of the buoyant part of the flux tube. The above ensures that the middle part of the flux tube will be buoyant, while the flanks will not. Thus the flux tube will adopt an $\Omega$-loop shape.
We use $\lambda=0.9$~Mm.

For PI we need to follow the same steps, but also take into account that changes in $\rho_i$ result in changes in the neutral fraction and therefore the reduced mass in the interior of the tube. $P_{exc}$ will be the same. By setting $P_i=P_0-P_{exc}$ and $T_i=T_0$, we get when the flux tube is in pressure balance and in thermal equilibrium with the background, the density in the interior of the flux tube is:
\begin{align}
    \rho_{i, teq} = \rho_0\frac{\mu_{mi}}{\mu_{m0}}\left( 1- P_{exc}/P_0 \right),
\end{align}
where $\mu_{mi}$ and $\mu_{m0}$ are the reduced mass of the interior of the flux tube and of the non-magnetized background. To make the flux tube adopt an $\Omega$-loop shape as before, we use a function that sets the density at the middle part of the flux tube to $\rho_i=\rho_{i, teq}$ and at the flanks to $\rho_i=\rho_0$:
\begin{align}
    \Delta\rho= - \rho_0 \left[ 1 - \frac{\mu_{mi}}{\mu_{m0}}\left( 1- P_{exc}/P_0 \right)\right] e^{-y^2/\lambda^2}.
    \label{eq:density_deficit_pi}
\end{align}
Indeed, notice that for $y=0$, $\rho_i=\rho_{i, teq}$ and that for $y\gg\lambda$, $\rho_i=\rho_0$. Also, for $\mu_{mi}=\mu_{m0}$ (i.e. non PI effects), we recover Eq.~\ref{eq:density_deficit}.

To calculate $\rho_i$ and $\mu_i$ , we numerically solve Eq.~\ref{eq:density_deficit_pi} in the following manner:
\begin{enumerate}
    \item First, $P_{exc}$ is calculated and $P_i=P_0-P_{exc}$ is set.
    \item Then, $\mu_{m0}$ of the background is found.
    \item An initial guess for $\mu_{mi}$ is assumed to be $\mu_{mi}=\mu_{m0}$.
    \item Then, $\rho_i$ is calculated from Eq.~\ref{eq:density_deficit_pi}.
    \item The internal energy $\epsilon_i$ is calculated from Eq.~\ref{eq:specific_energy}. Then, the temperature $T_i$ is calculated from Eq.~\ref{eq:ideal_gas_law}
    \item Based on the current values of temperature and density, the new $\xi_{ni}$ and $\mu_{mi}$ are computed.
    \item Steps 4-6 are repeated until $\xi_{ni}$ converges to a value.
\end{enumerate}
Using the above method, we derive the temperature, specific energy, density and gas pressure profile of the flux tube and the corresponding neutral fraction of the magnetized plasma.

\section{Results}
\subsection{Magnetic flux emergence at the solar interior}
\begin{figure}[h]
\centering
\includegraphics[width=1.1\columnwidth]{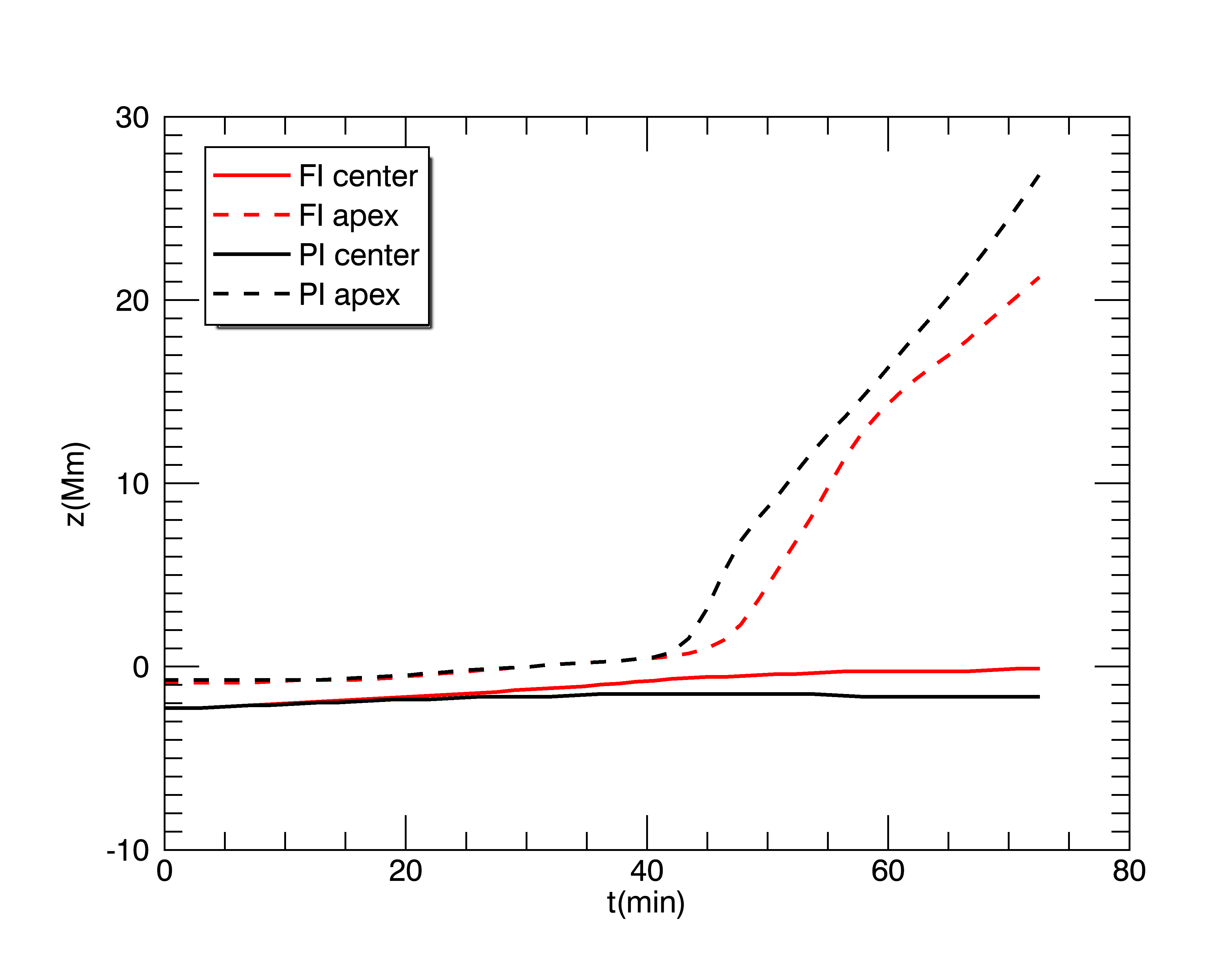}
\caption{Height-time profile of the apex and center of the tubes in both simulations.}
\label{fig:Apex}
\end{figure}

To study the emergence of magnetic flux below the photosphere, we first follow the rise of the centre and the apex of the tube in both simulations. We define the centre 
of the tube as the location at the vertical xz-midplane, where $B_{y}$ is maximum and $B_{x}$ changes its sign across the tube, along height. For simplicity, we define 
the apex of the tube to be at the outskirts of the magnetic field, where the magnetic field strength decreases to a value of about $0.001\times B_0$. Figure \ref{fig:Apex} shows 
the time evolution of the emergence in both simulations.
\par We find that, in both simulations, the tube's axis emerges in a similar fashion until about $t=25$ minutes (figure \ref{fig:Apex}). However, after $t=25$ minutes, we find 
that the two flux tubes emerge in a different manner.
The axis of the tube in FI continues to rise until it reaches the photosphere, while in the other case the axis of the tube stays at around $2 Mm$ below the 
photosphere until the end of the simulation.
A similar result was reported by \cite{Leake_etal2013b}, who suggested that the reason for this difference is the ionization-recombination effect, induced 
by the ion-neutral collisions, at the upper part of the solar interior. In Figure \ref{fig:Apex}, we 
also find that the expansion of the magnetic field above the photosphere occurs earlier in the PI simulation. This is because, the inclusion of the ion-neutral collisions and the 
increased dissipation allows the magnetic field to slip through the plasma during emergence and, thus, it doesn't carry within it so dense plasma as in the FI case. Thus, 
the apex of the field is less heavy in the PI case and it can expand earlier.

\par
\begin{figure*}[h]
\includegraphics[width=0.345\textwidth]{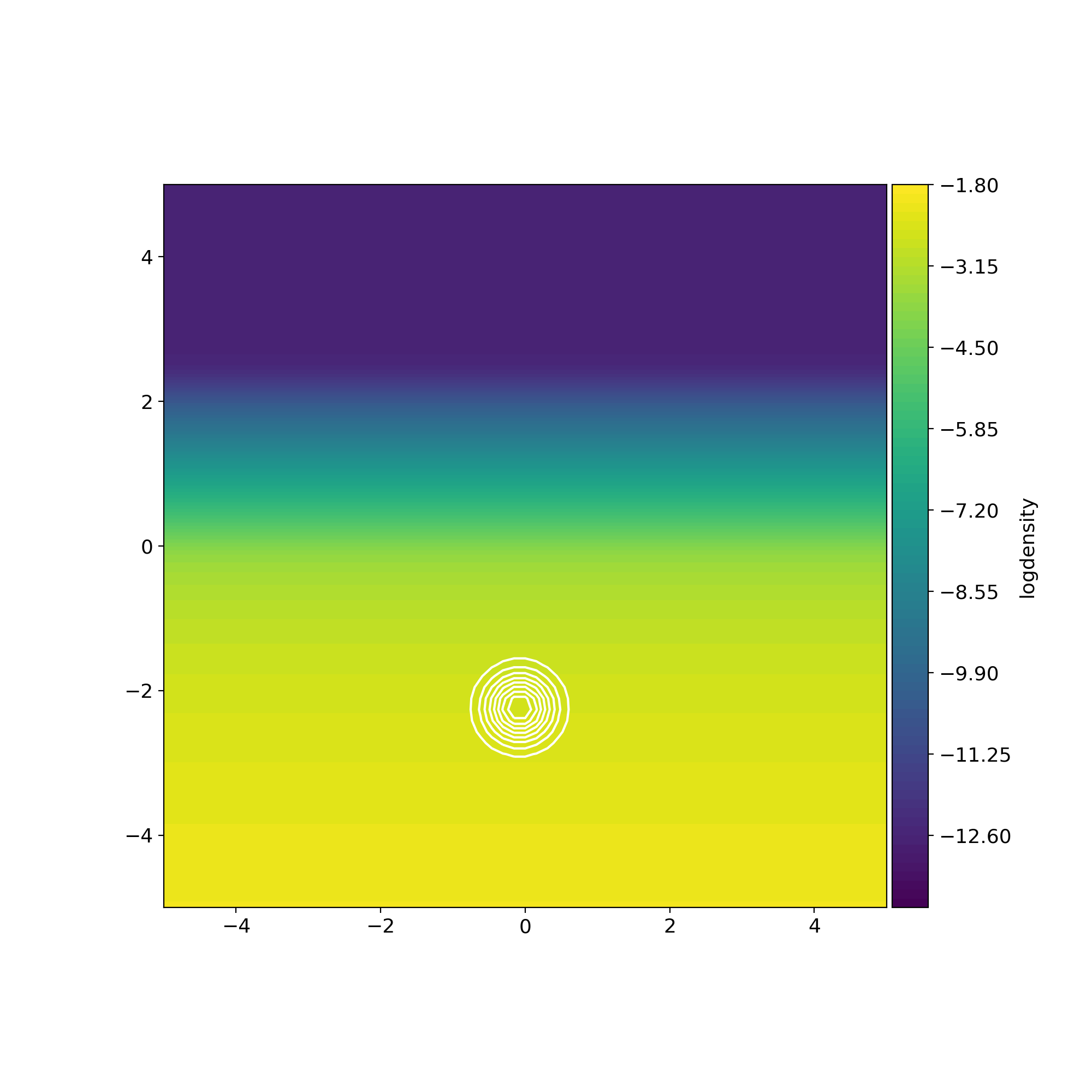}
\includegraphics[width=0.345\textwidth]{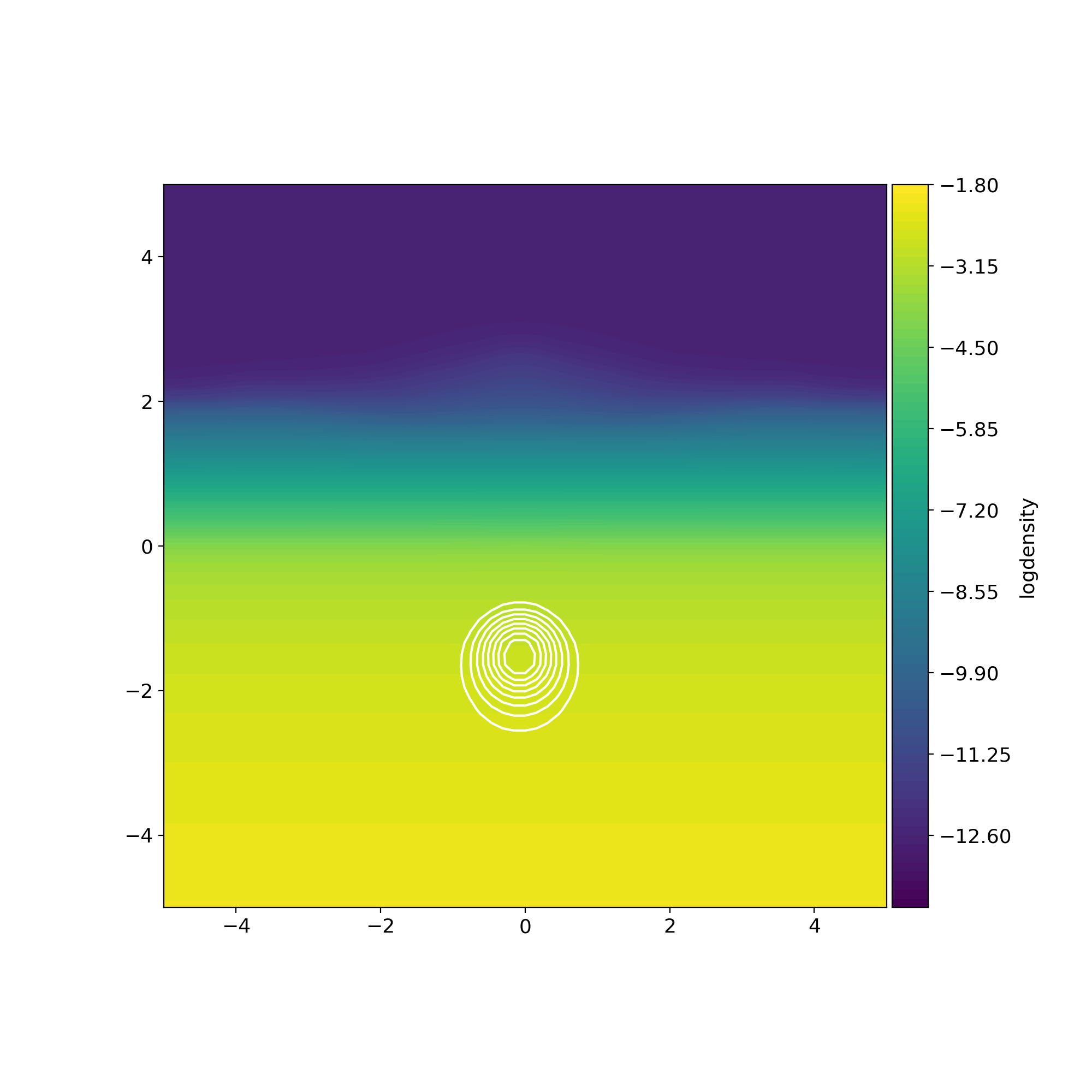}
\includegraphics[width=0.345\textwidth]{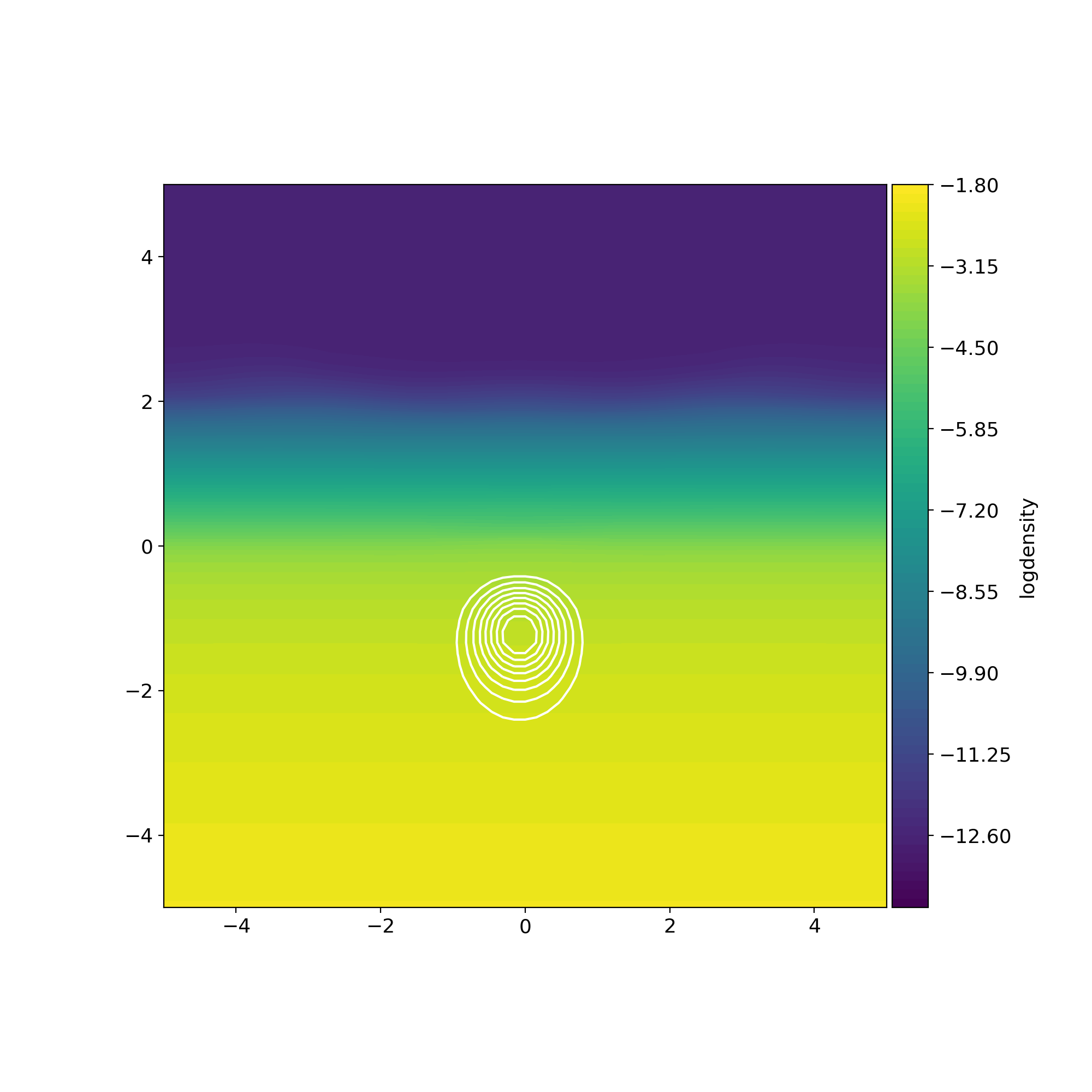}
\includegraphics[width=0.345\textwidth]{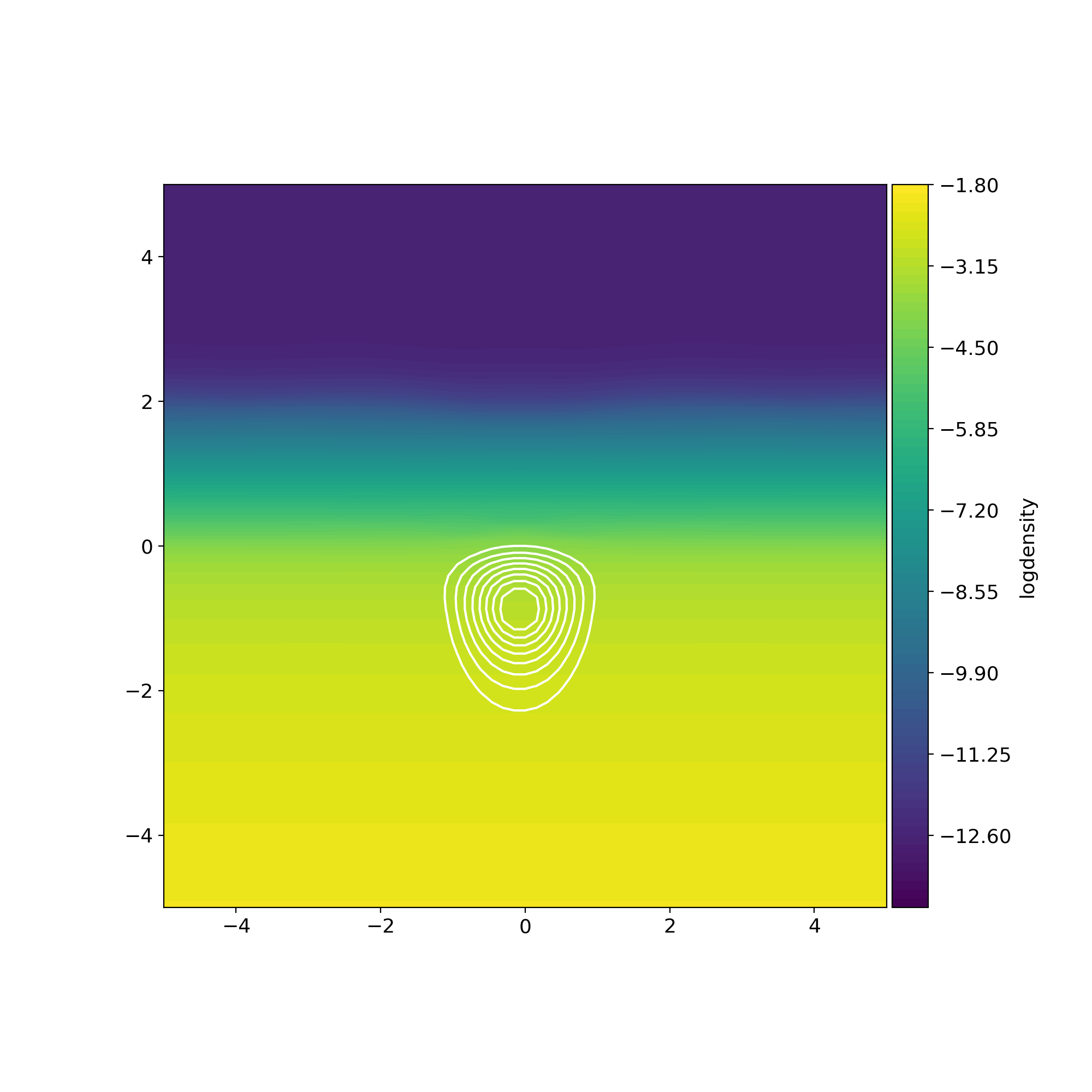}
\includegraphics[width=0.345\textwidth]{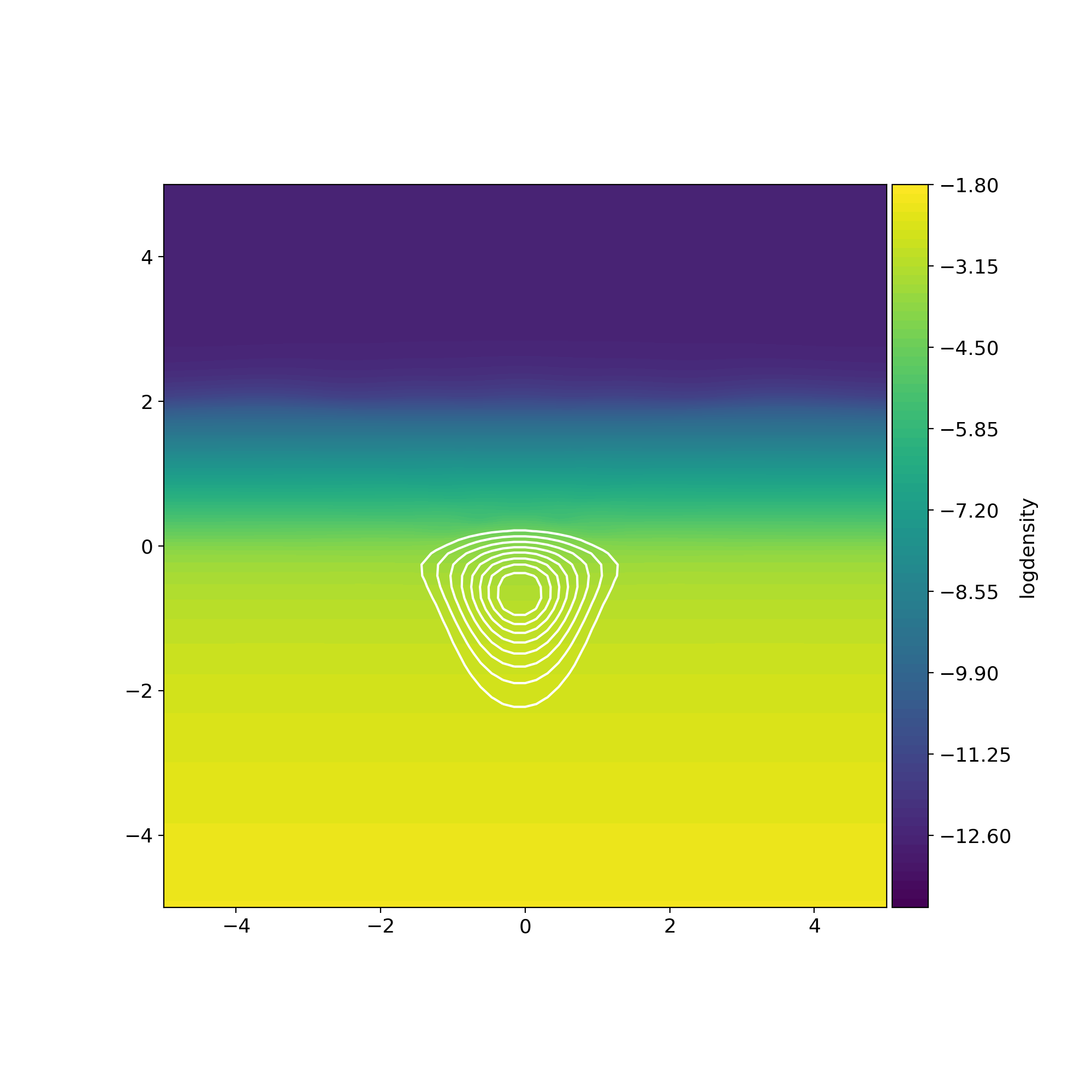}
\includegraphics[width=0.345\textwidth]{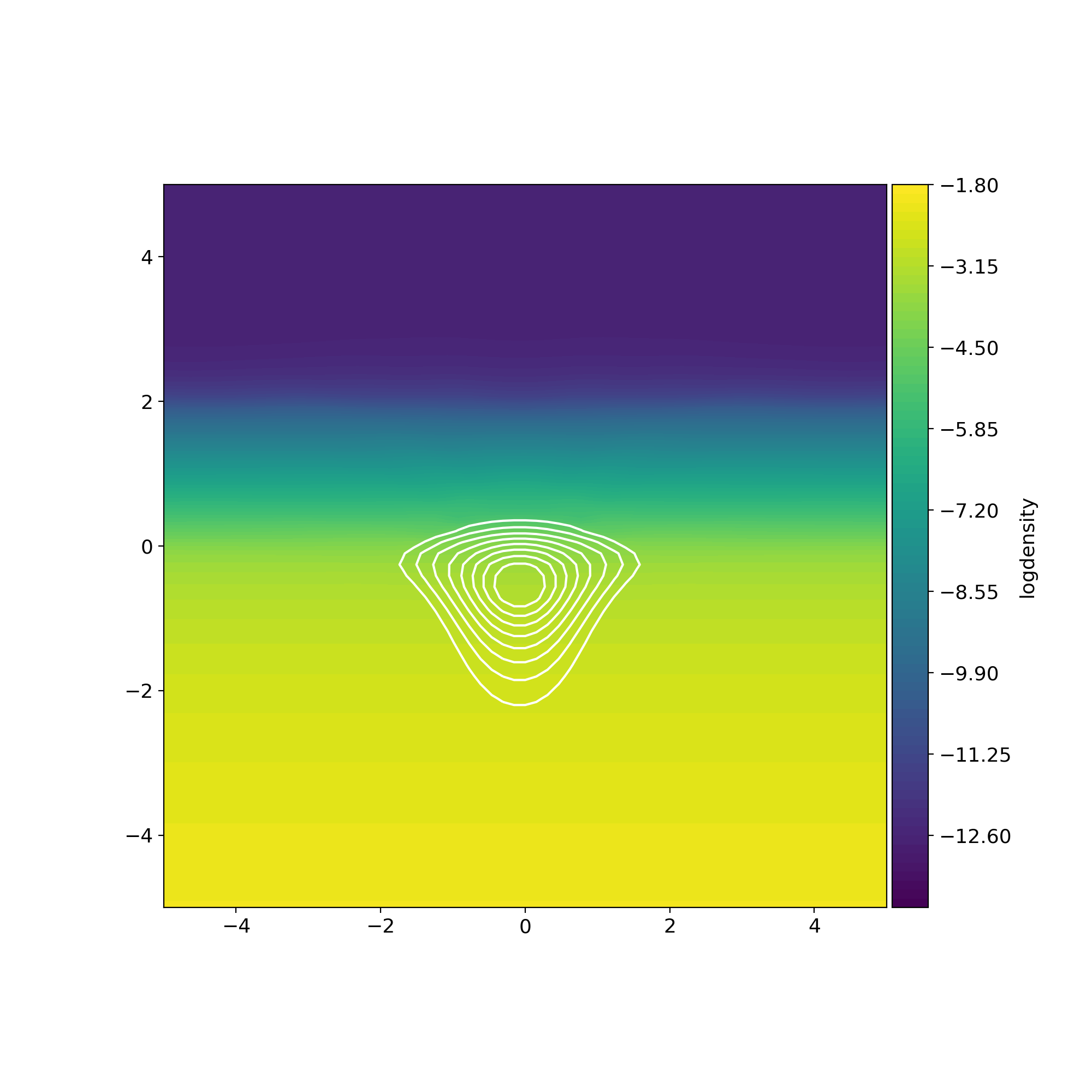}
\caption{The evolution of By on xz plane on the FI simulation from up left to bottom right at t=0,22,29,36,40,43 minutes. }
\label{Fig:ByFIP}
\end{figure*}

\begin{figure*}
\includegraphics[width=0.345\textwidth]{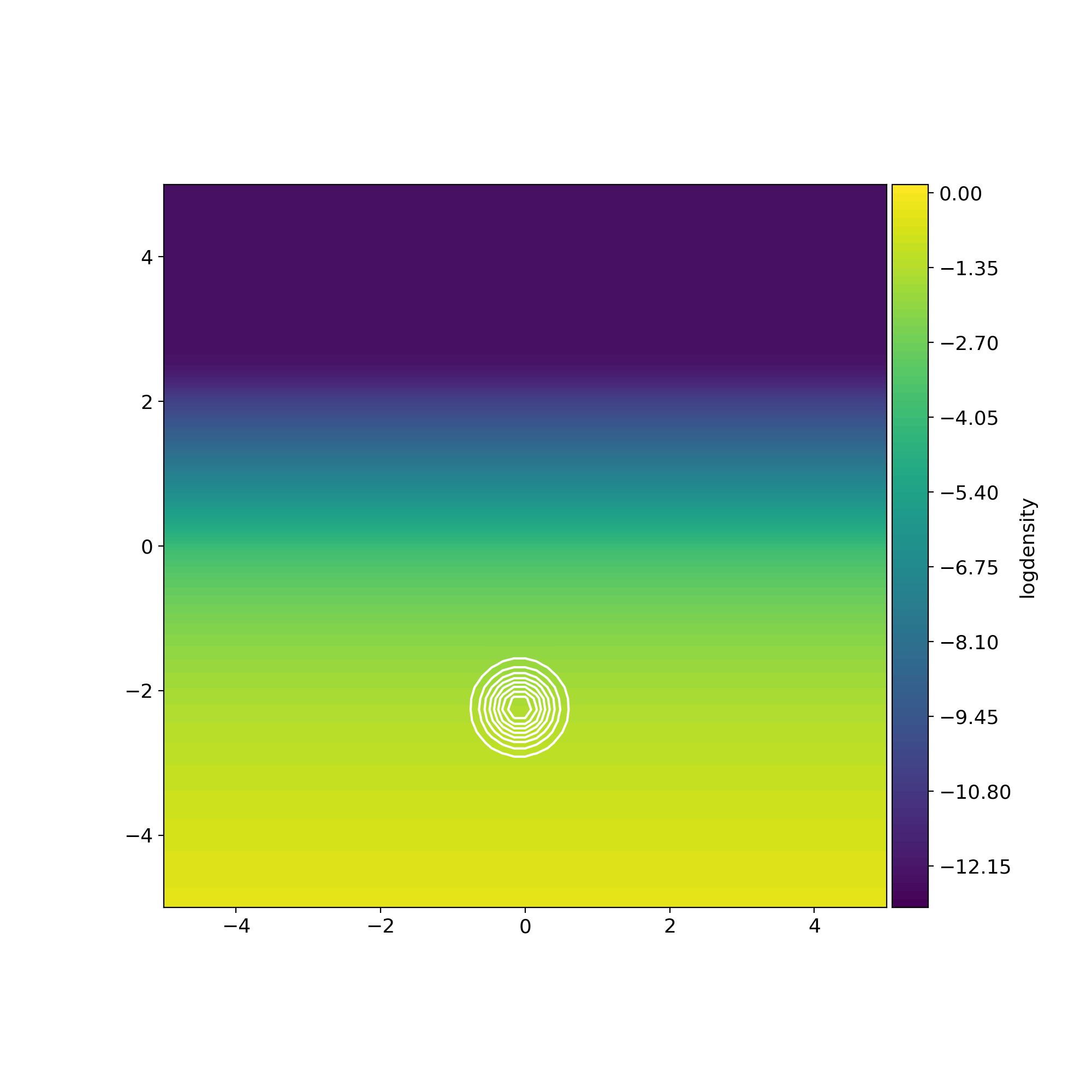}
\includegraphics[width=0.345\textwidth]{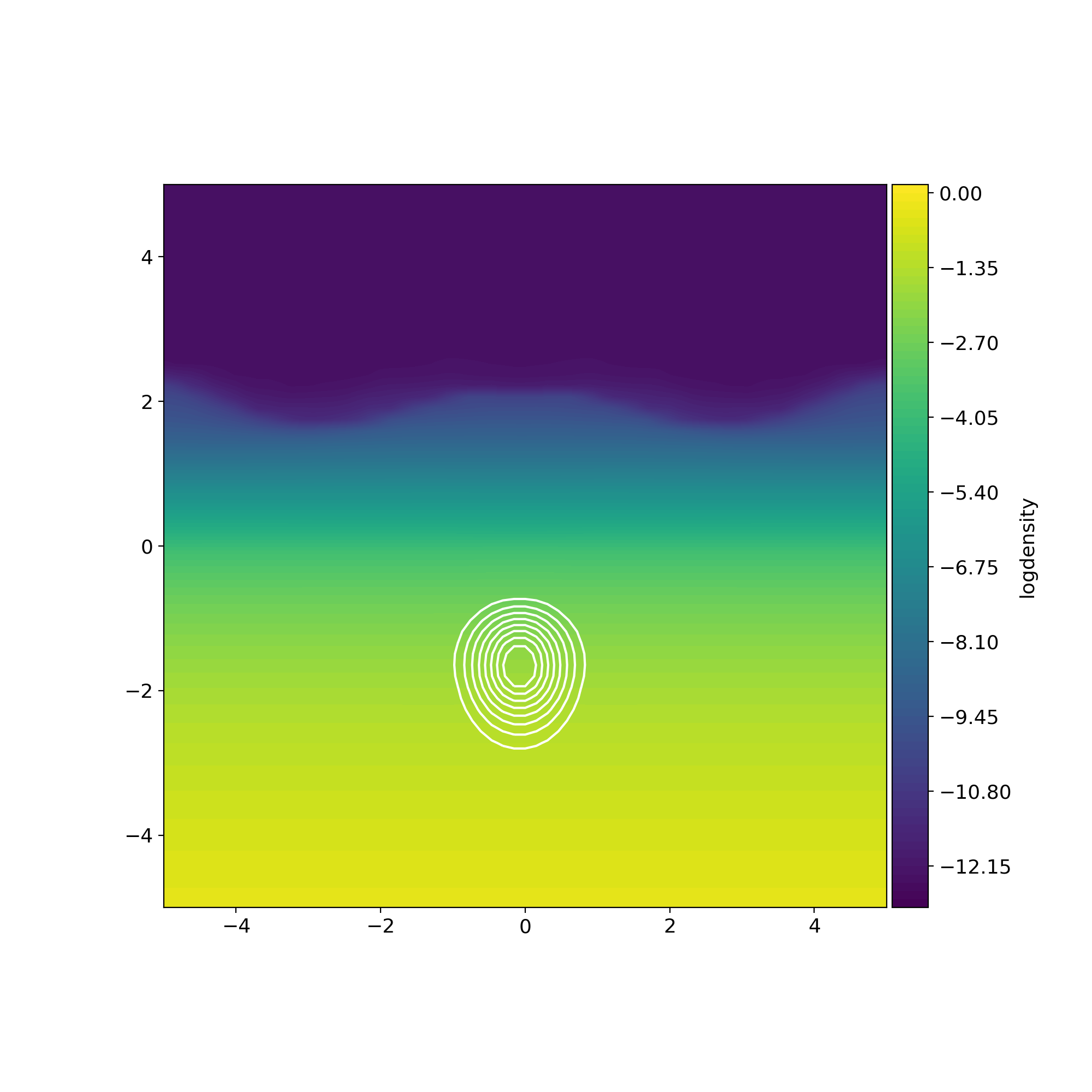}
\includegraphics[width=0.345\textwidth]{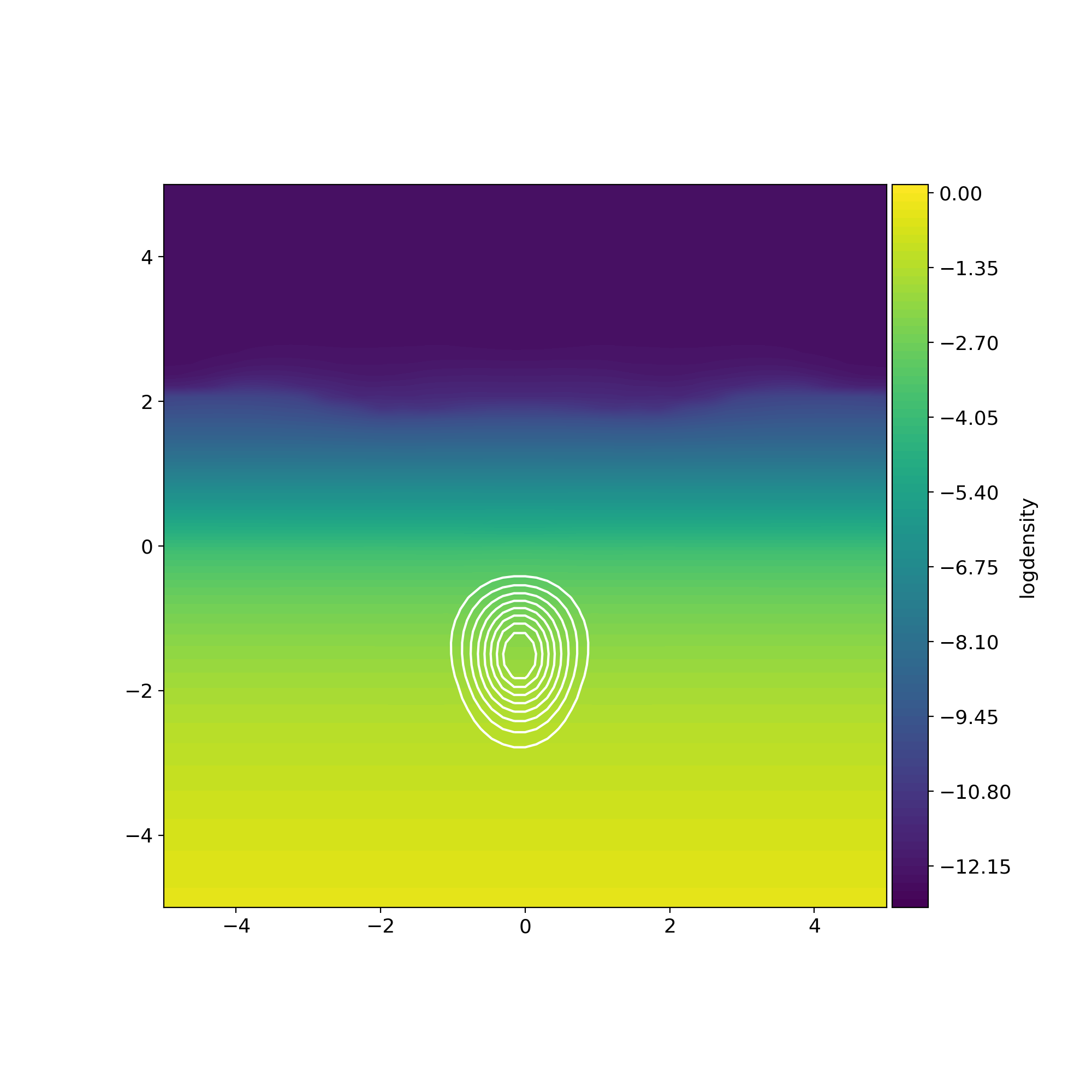}
\includegraphics[width=0.345\textwidth]{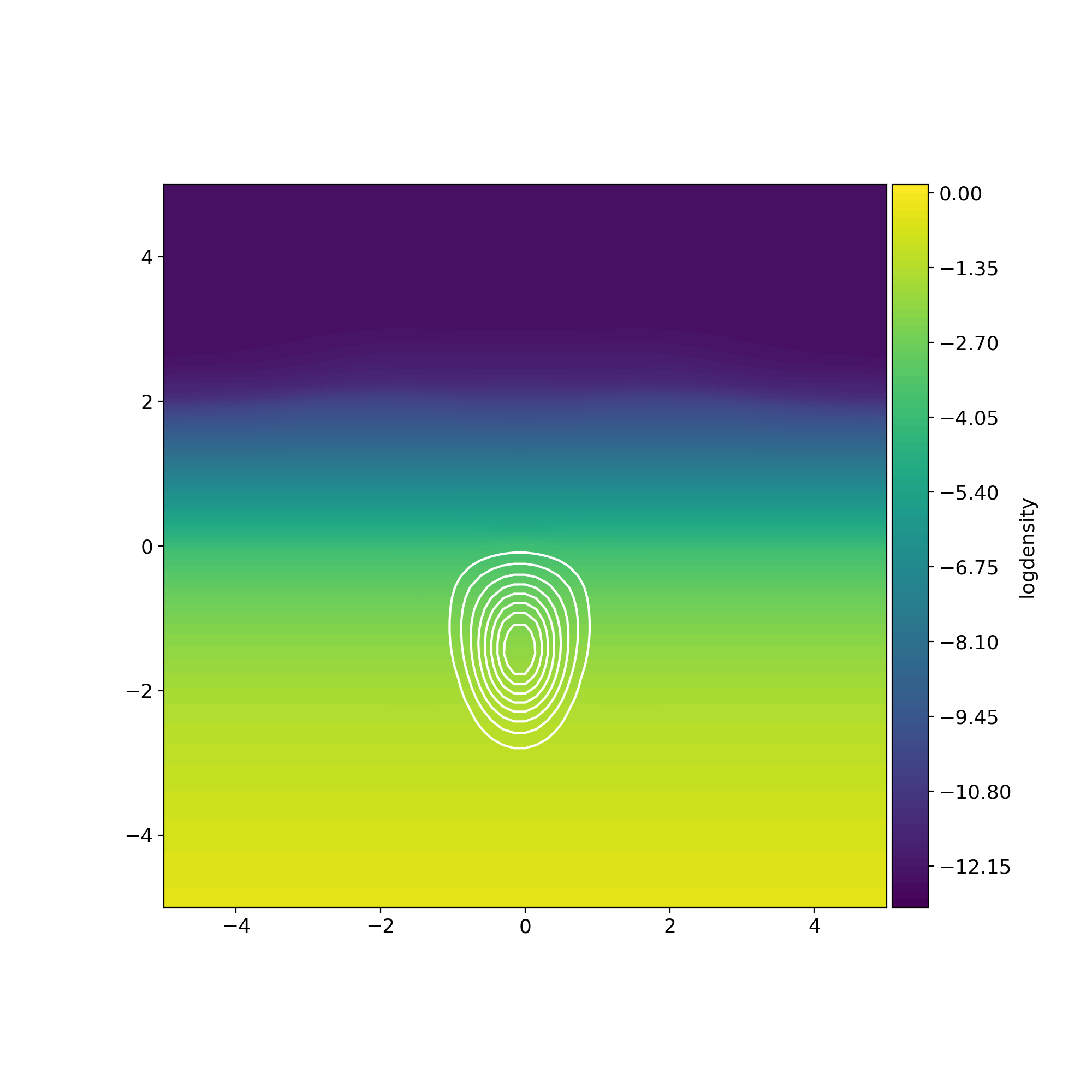}
\includegraphics[width=0.345\textwidth]{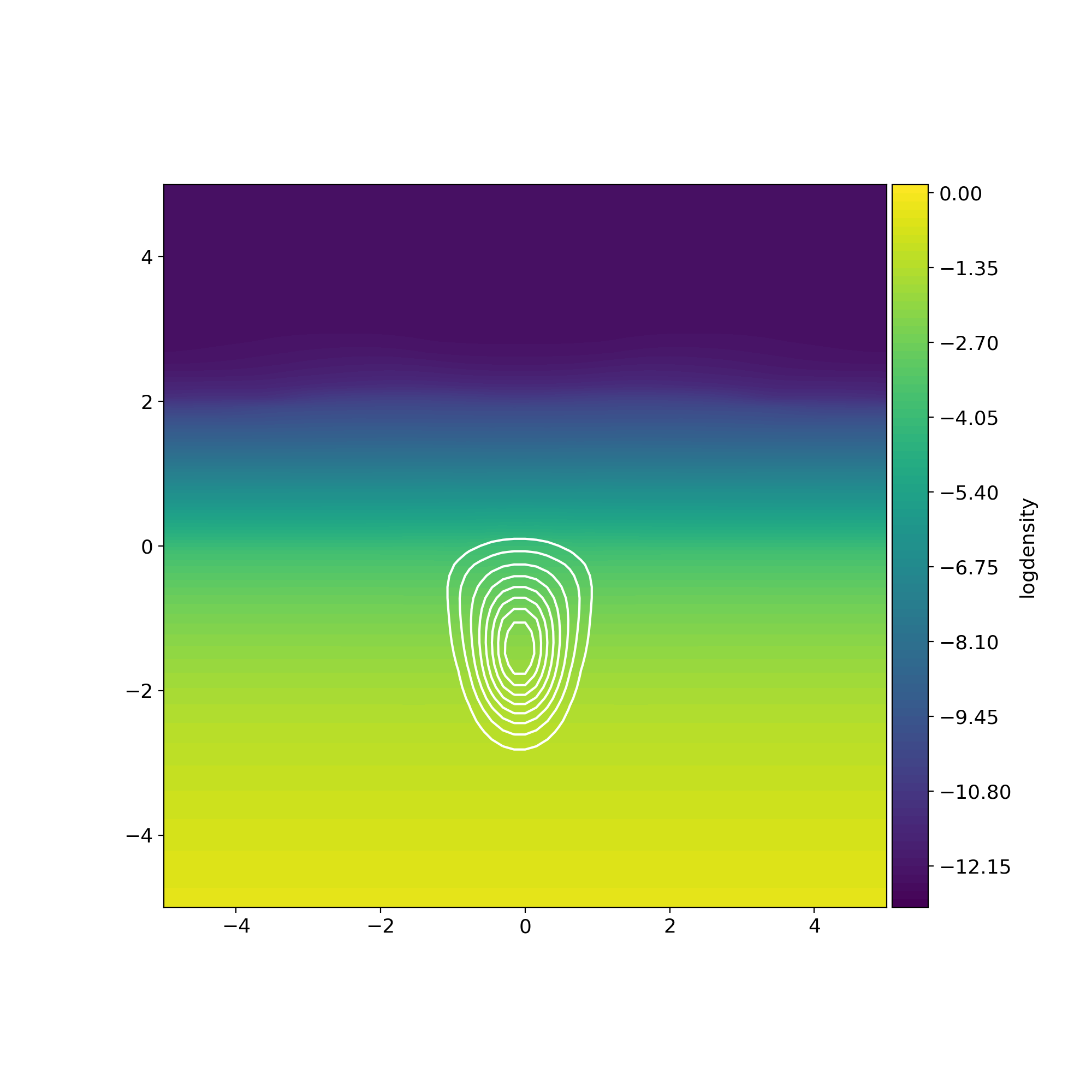}
\includegraphics[width=0.345\textwidth]{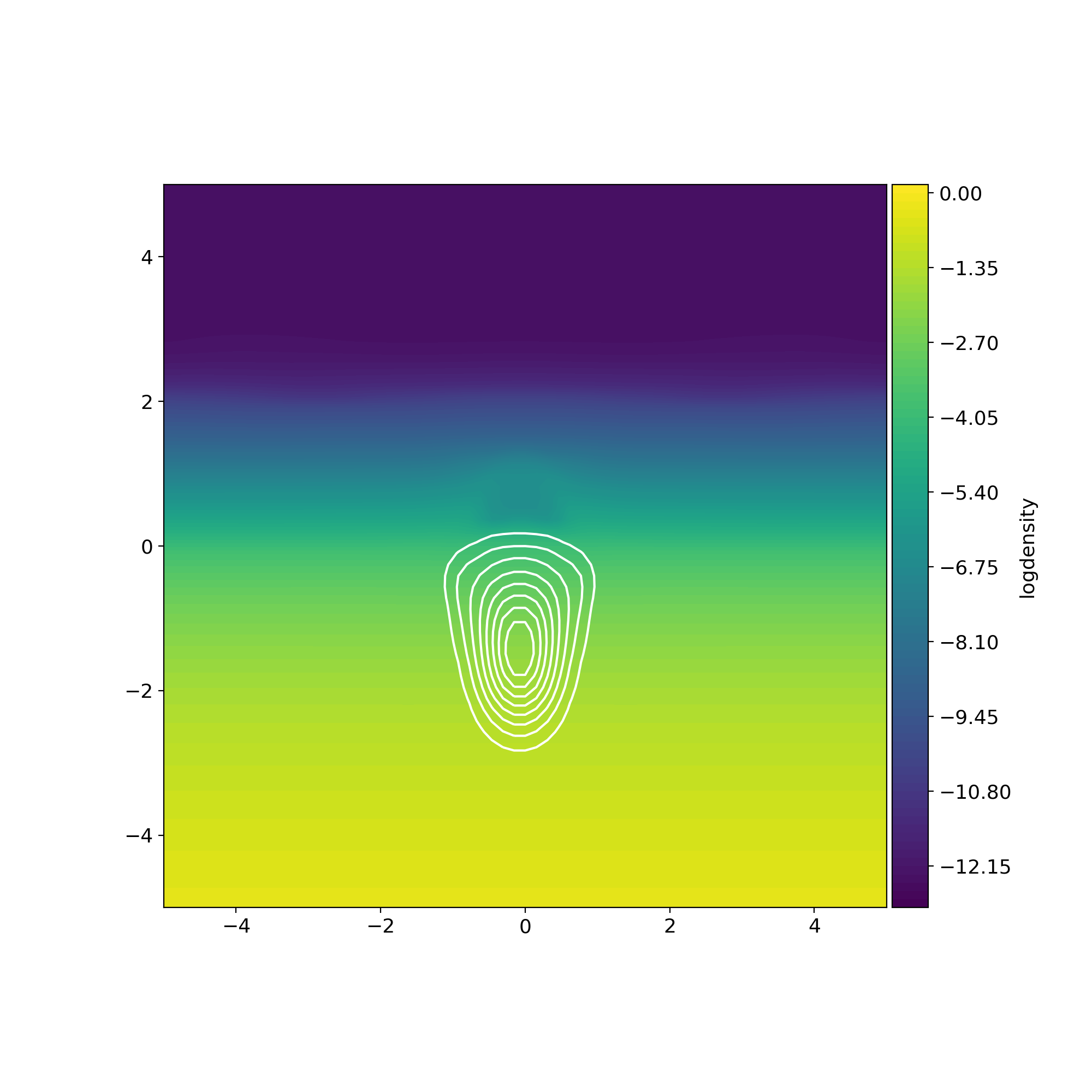}
\caption{The evolution of By on xz plane on the PI simulation from up left to bottom right at t=0,22,29,36,40,43 minutes. }
\label{Fig:ByPIP}
\end{figure*}

 A similar effect is apparent also below the photosphere. Figure \ref{Fig:ByFIP} and Figure \ref{Fig:ByPIP} show the emergence of the magnetic 
flux tube in the solar interior for the two cases. 
The contours show the cross-section of the tube at the vertical xz-midplane and they correspond to values of $B_{y}$ greater than $3\times10^{-3}T$. In the FI case (figure \ref{Fig:ByFIP}), the emergence of the field follows 
a typical behaviour, as this has been shown in many similar magnetic flux emergence experiments in the past. Namely, the tube rises vertically upwards and when it 
reaches the photosphere, it slows down and expands horizontally before it becomes unstable and emerges above the photosphere. In the PI case (figure \ref{Fig:ByPIP}) 
the shape of the cross-section of the tube is different, especially its top part and while it approaches the photosphere. We find that it is less flat and it does not 
expand horizontally as in the FI case, adopting an overall spindle-like shape. For instance, at $t=22$, $t=33$ and $t=43$ minutes, we observe 
a more vertical than horizontal stretching/expansion. We anticipate 
that it is the perpendicular resistivity that dissipates the cross-field currents and this permits the magnetic field to “slip" through the plasma, giving this 
characteristic shape to the emerging field. As a consequence, the emerging bipolar region at the photosphere adopts a different shape as we show in the next section.

\subsection{Magnetic flux emergence at the photosphere}
\par
\begin{figure*}
\includegraphics[width=0.245\textwidth]{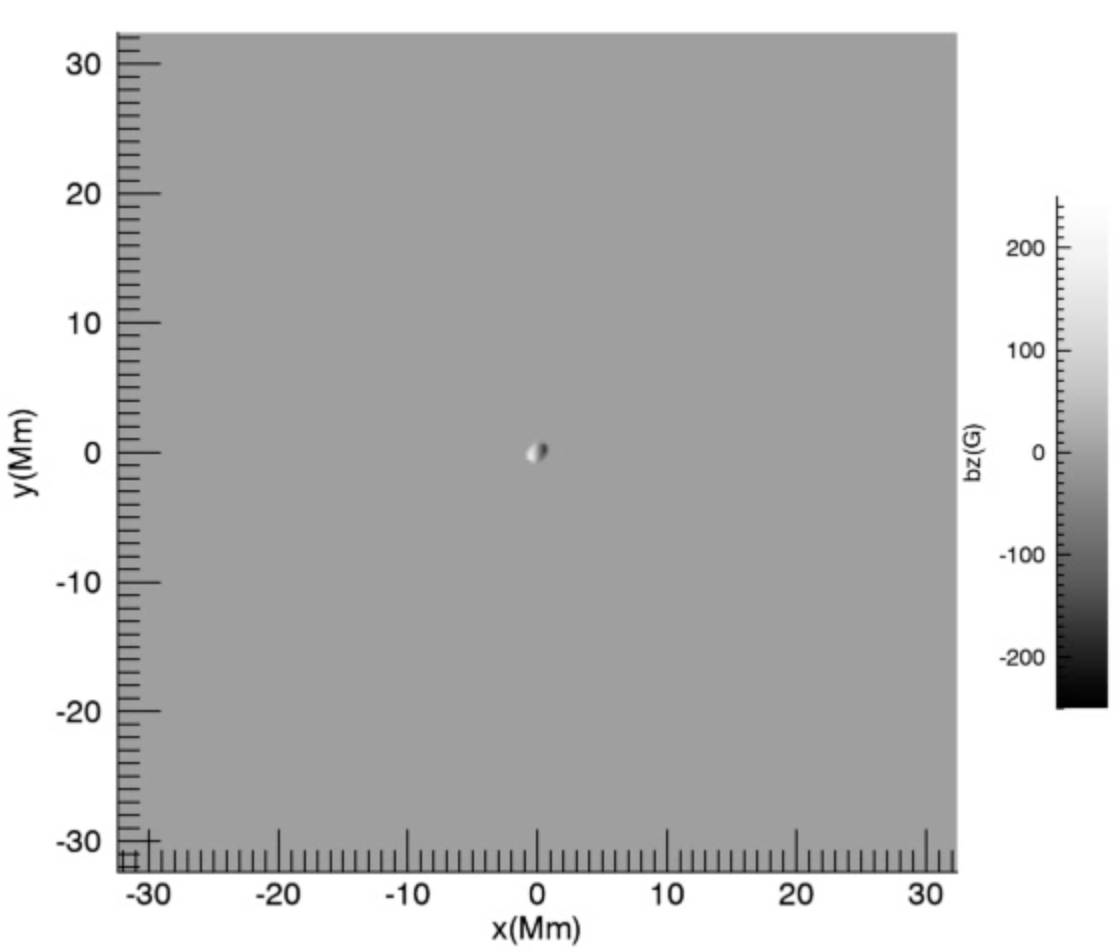}
\includegraphics[width=0.245\textwidth]{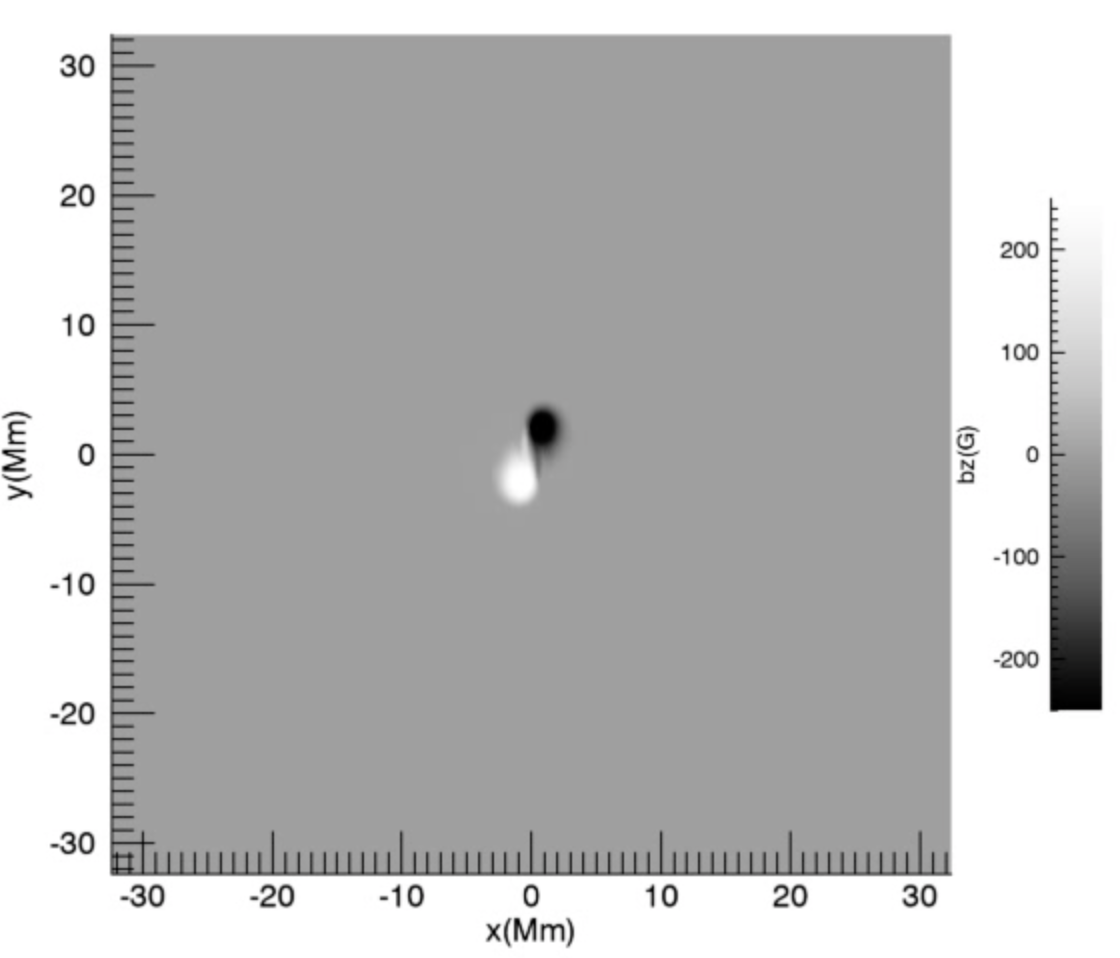}
\includegraphics[width=0.245\textwidth]{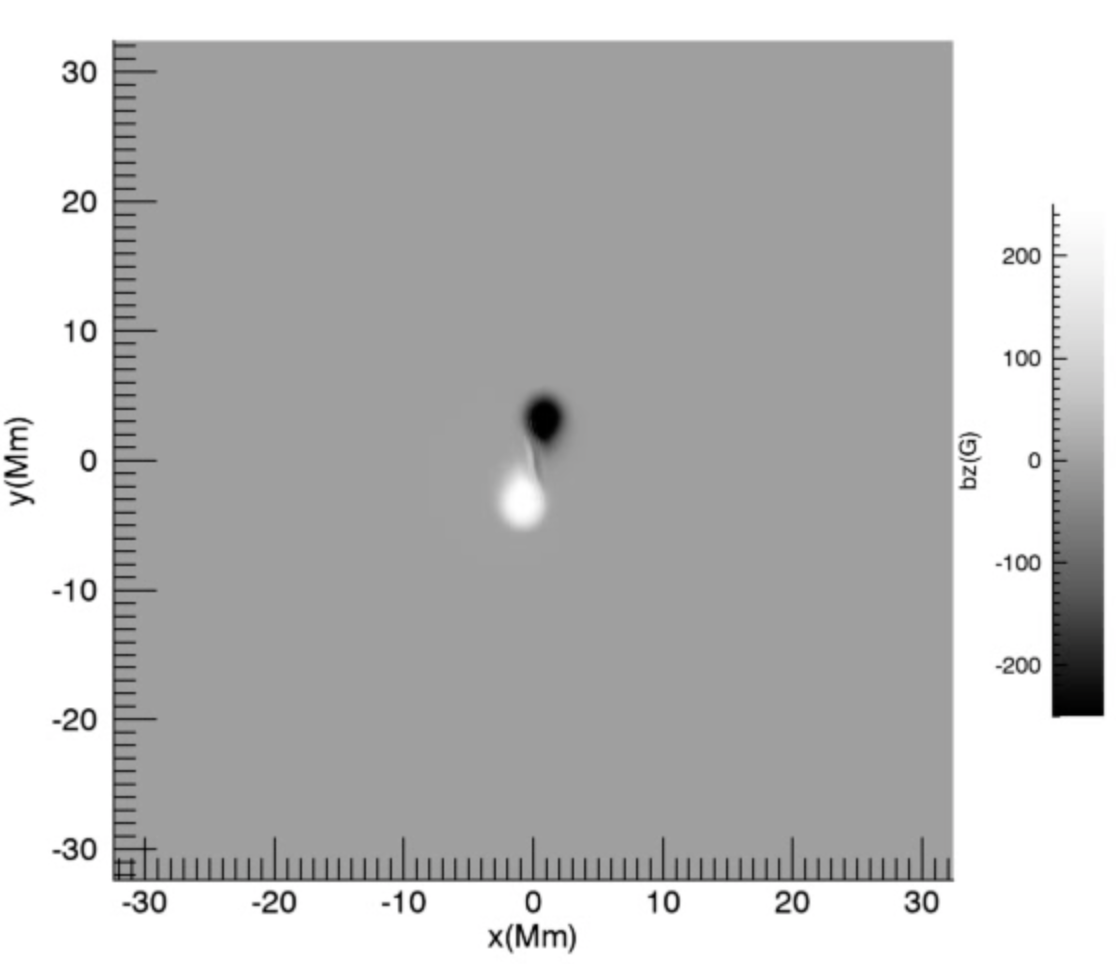}
\includegraphics[width=0.245\textwidth]{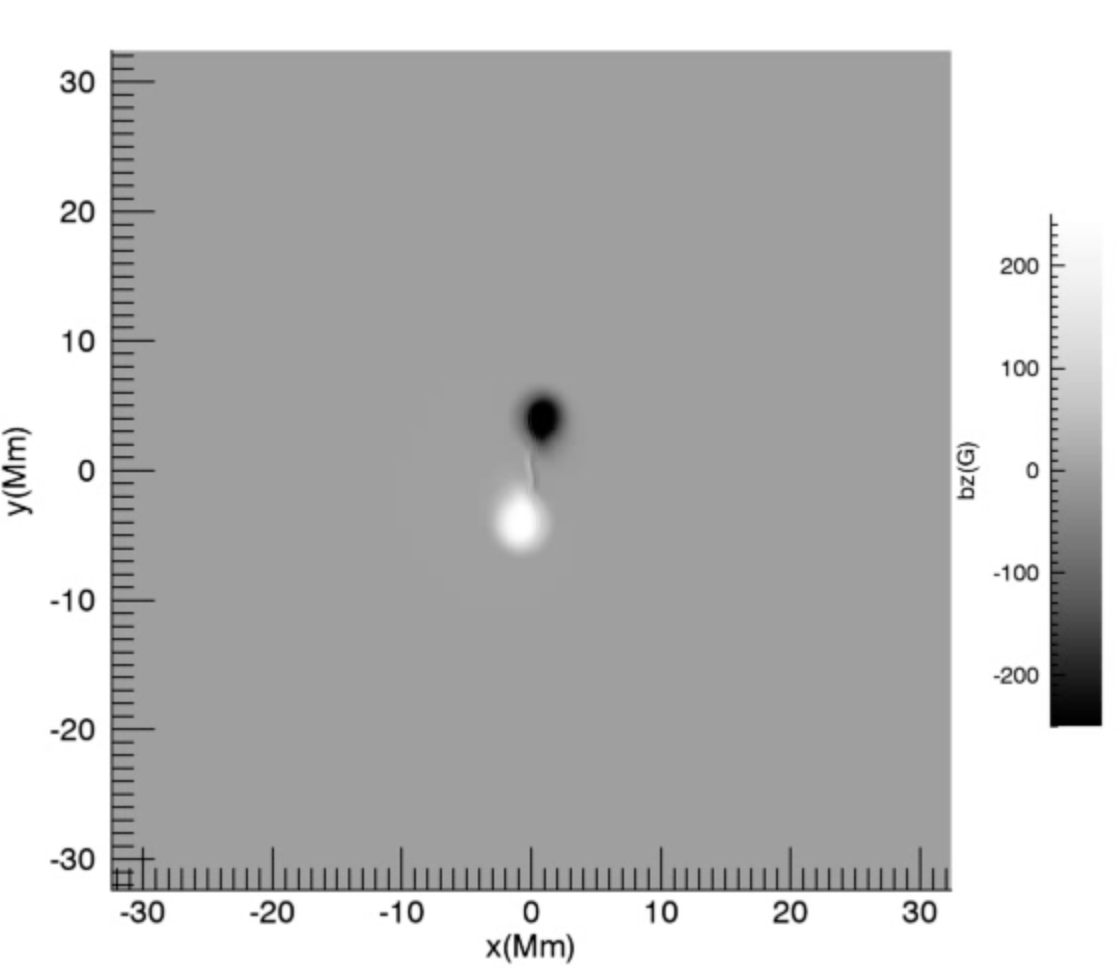}
\includegraphics[width=0.245\textwidth]{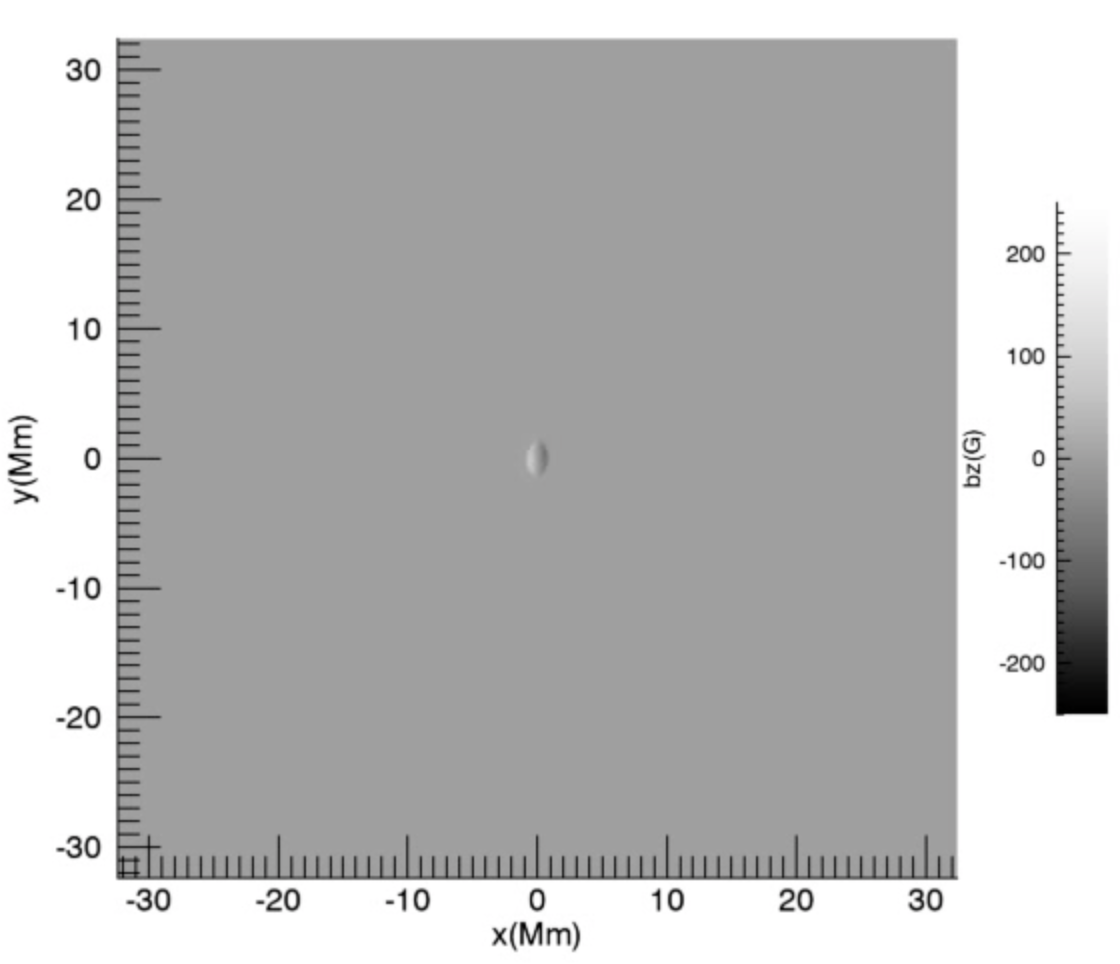}
\includegraphics[width=0.245\textwidth]{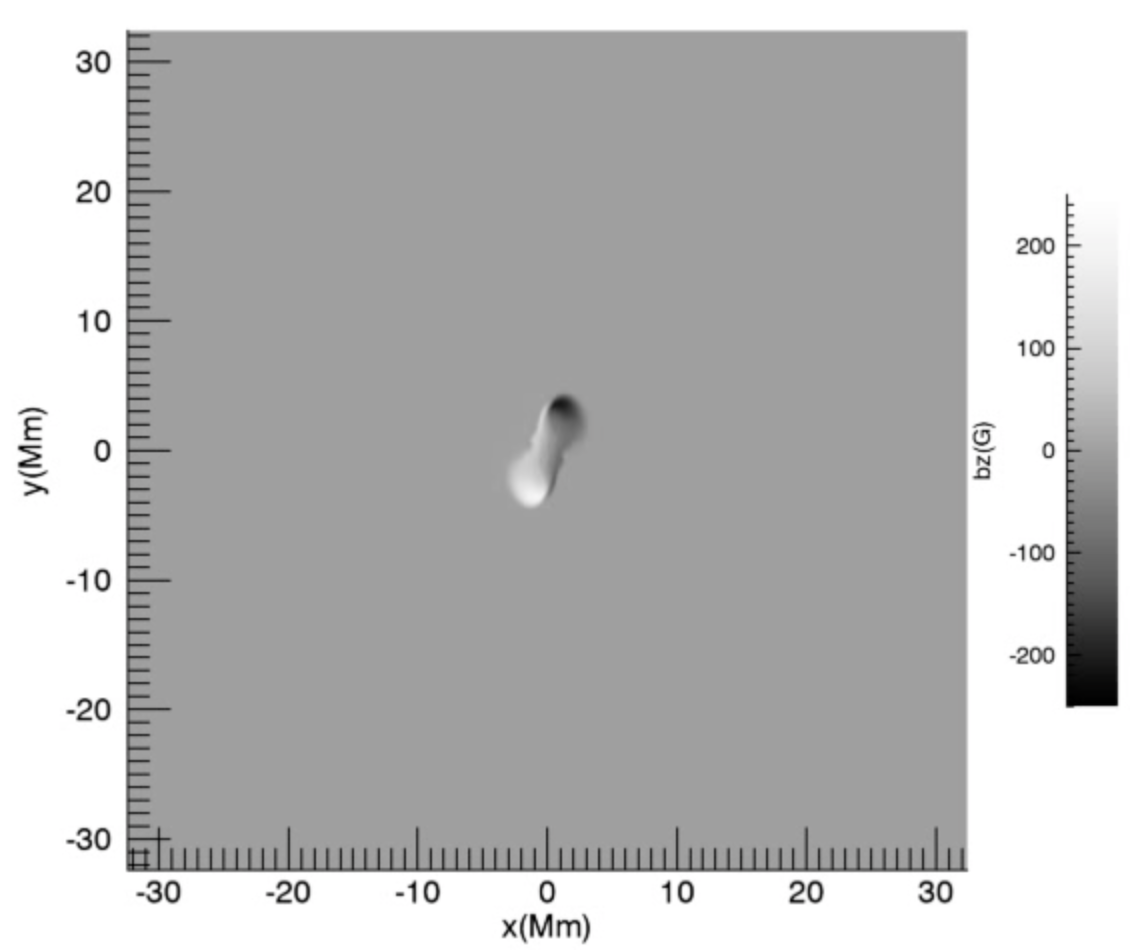}
\includegraphics[width=0.245\textwidth]{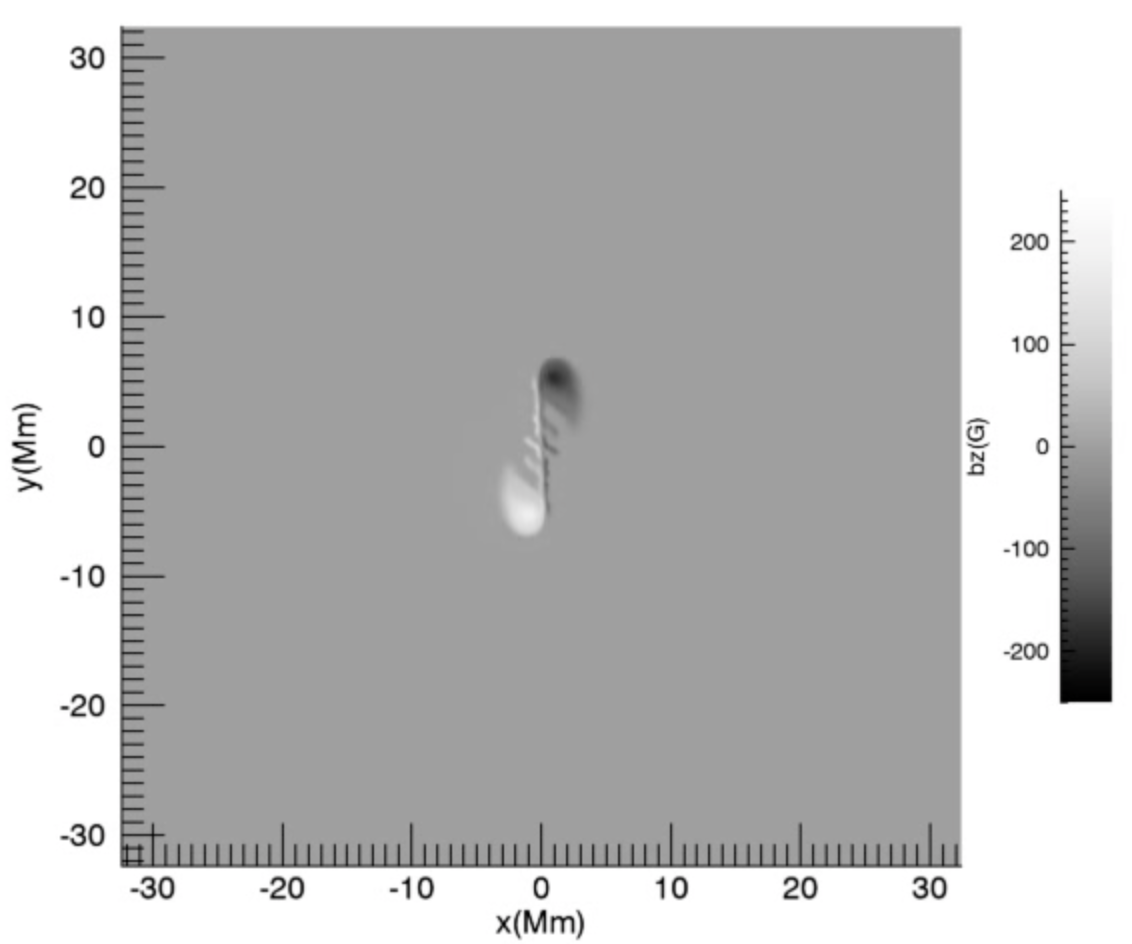}
\includegraphics[width=0.240\textwidth]{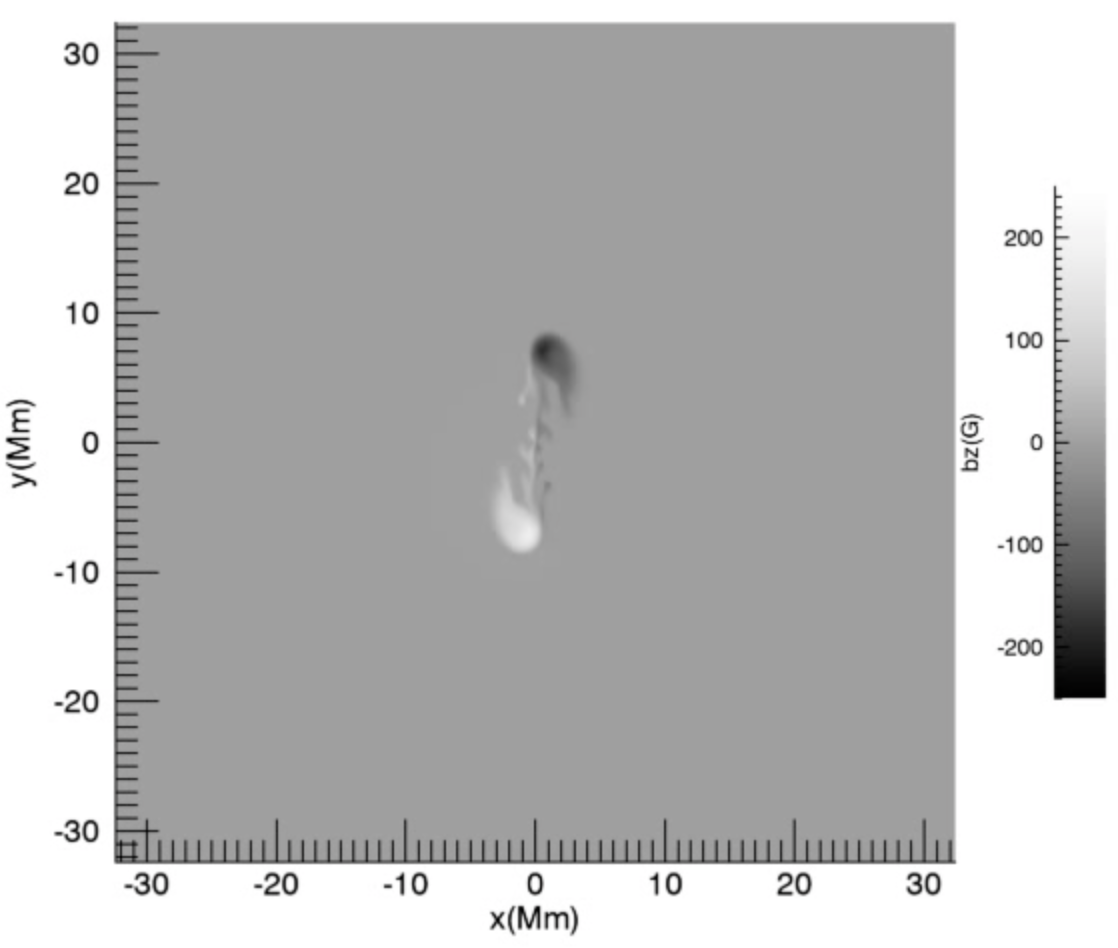}
\caption{The evolution of  bipolar region on PI simulation(first row) and FI simulation (second row) from left to right 43,72,115 and 159 minutes. }
\label{Fig:BPFIP}
\end{figure*}

\begin{figure}[]
\centering
\includegraphics[width=\columnwidth]{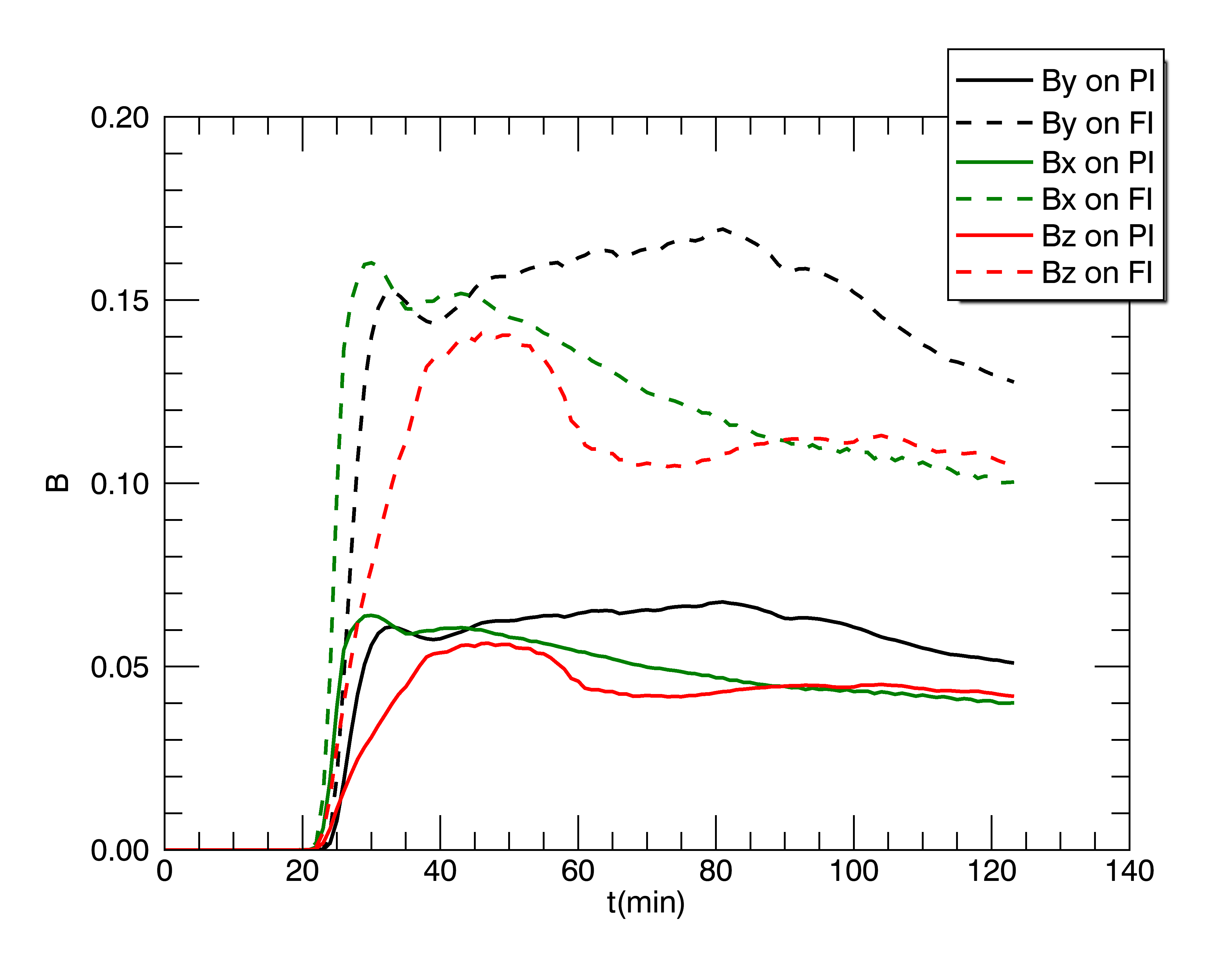}
\caption{Time evolution of the normalized B components at the base of the photosphere.}
\label{fig:components}
\end{figure}

 Figure \ref{Fig:BPFIP} shows the vertical component of the magnetic field at the photosphere at different times.
In general, the time evolution shows the typical behaviour of an emerging bipolar region. First, the two
opposite polarities appear along the x-axis, almost perpendicular to the direction of the original
axis of the tube due to the strong twist and then they move apart along the y-axis. In between them, a strong polarity inversion line (PIL) is formed. There are some profound differences in the appearance and evolution of the emerging region in the two cases. In the PI case, the two polarities appear to adopt a more circular and compact shape. In a qualitative manner, this result is consistent with the shape of the emerging cross-section of the tube in Figure \ref{Fig:BPFIP} (first row), where the magnetic field contours appear more vertical at the intersection with the photosphere in the PI case. Also, Figure \ref{Fig:BPFIP} (second row) shows that in the FI case the magnetic tails, on the two sides of the PIL, are more noticeable. To understand better the reason(s) for the manifestation of these differences we study the evolution of the individual components of the magnetic field at the photosphere. More precisely, in Figure \ref{fig:components}, we plot the maximum values of the components at the x=y=0 and at the base of the photosphere. We normalize them with the maximum value of the initial total magnetic field. Firstly, we notice that, overall, a larger percentage of the magnetic field components emerge to the photosphere in the FI case. Secondly, we find that in the PI case, the dominant component is Bz, which explains the more profound appearance of the two polarities at the photosphere in Figure \ref{Fig:BPFIP} (first row). In the FI case the dominant component is the axial component By, which is consistent with the height-time profile of the
axis of the tube in Figure \ref{fig:Apex}. The second strongest component in the FI case is the azimuthal component of the magnetic field, while it is the
weakest in the PI case. This explains why the magnetic tails are more noticeable in the FI case.


\begin{figure}[]
\centering
\includegraphics[width=\columnwidth]{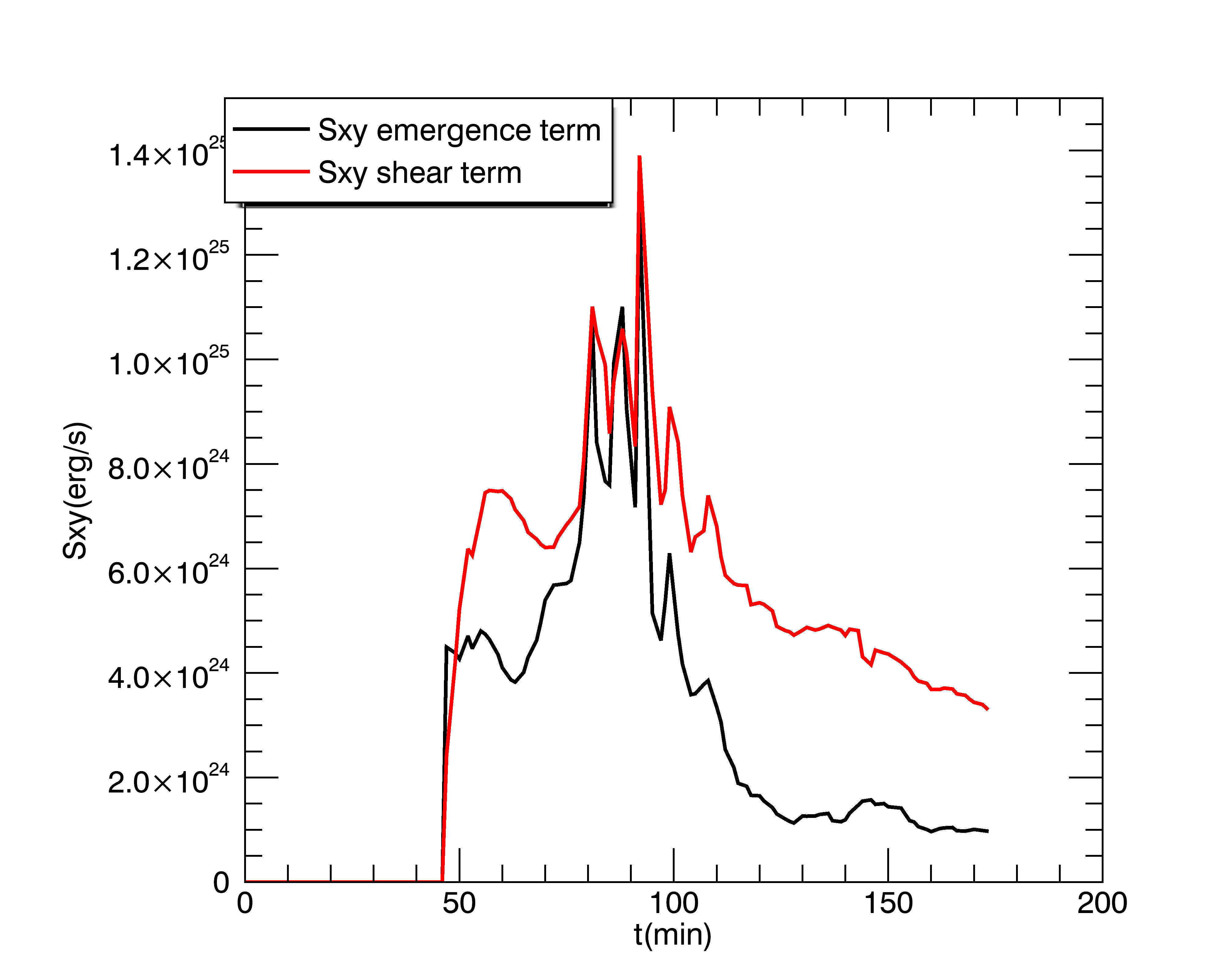}
\caption{Emergence and shear term of Poynting flux on FI simulation at the mid-photosphere.}
\label{fig:PoyntingFip}
\end{figure}
\begin{figure}[]
\centering
\includegraphics[width=\columnwidth]{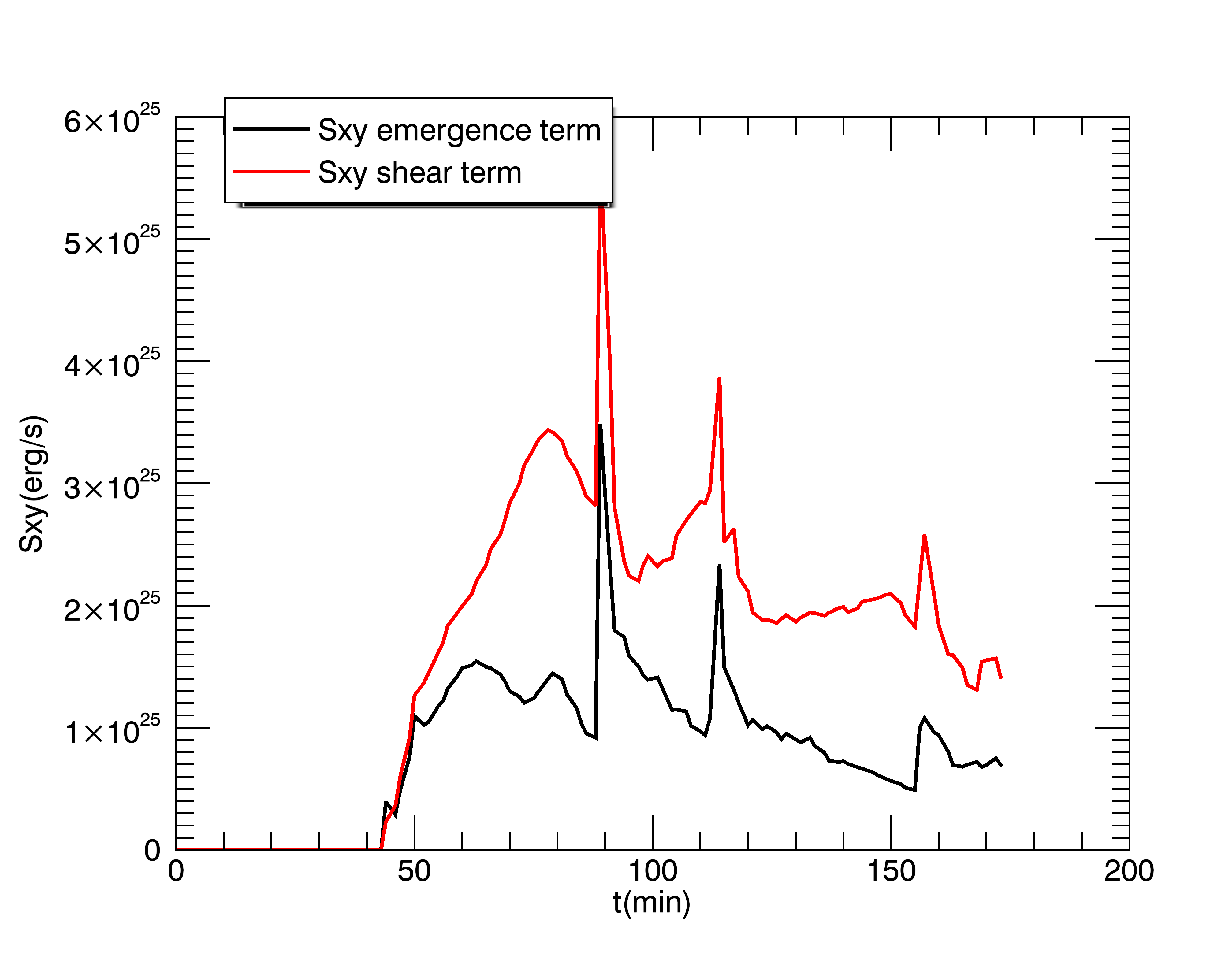}
\caption{Emergence and shear term of Poynting flux on PI simulation at the mid-photosphere.}
\label{fig:PoyntingPip}
\end{figure}

\begin{figure}[]
\centering
\includegraphics[width=0.445\textwidth]{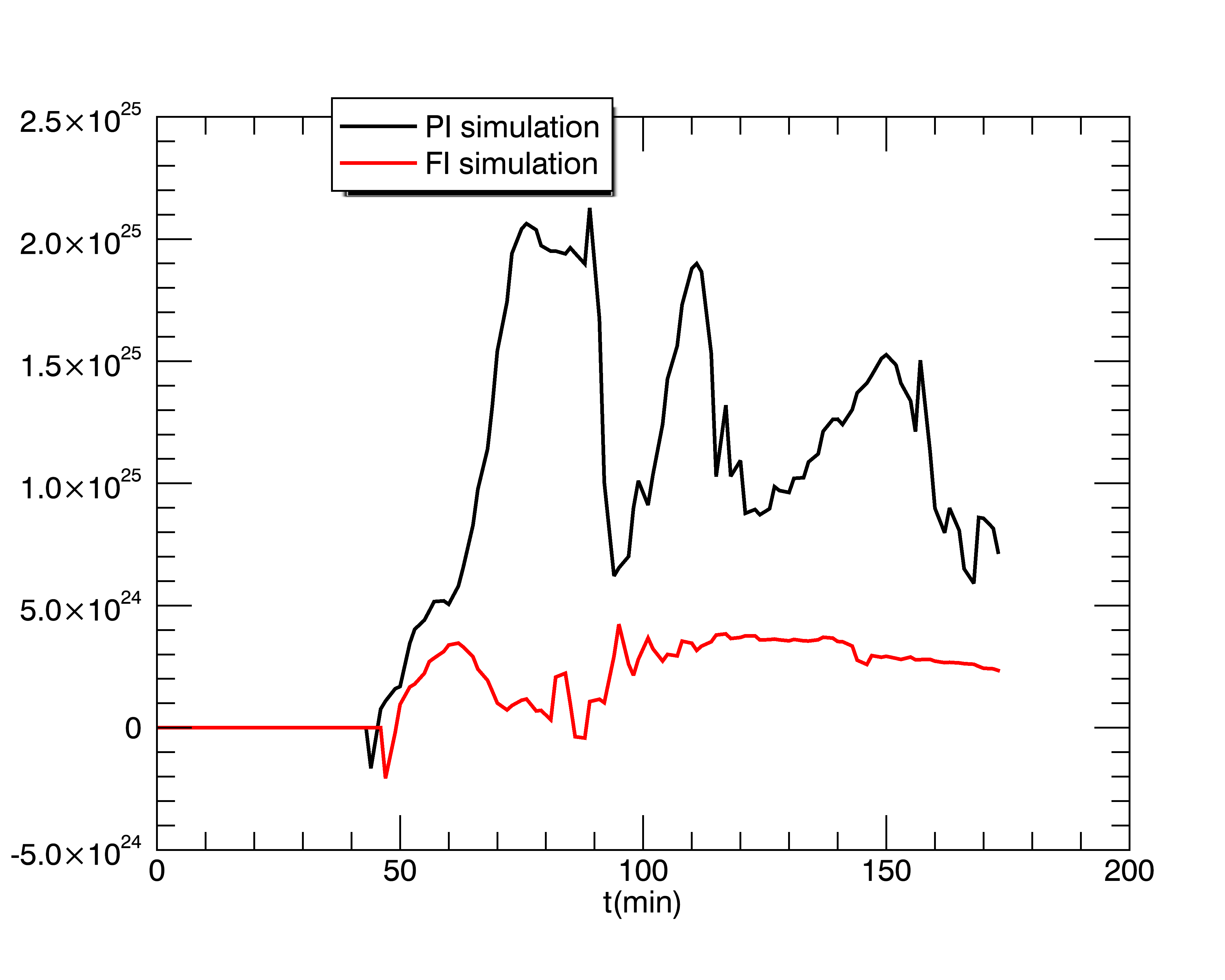}
\caption{The difference between the shear and emergence terms of the Poynting flux.}
\label{fig:difference}
\end{figure}

\subsection{Flux and energy evolution}
\par To study the relative importance between the emergence and shearing, we follow the time evolution of the Poynting flux at z=1.2Mm . More precisely, we have split the Poynting flux into two terms, the ``emergence" and the ``shear" term as shown below:
\begin{align}
    S_{xy}{emergence}&=+\frac{1}{4\pi}\int_{x} \int_{y} v_z(B_{x}^2+B_{y}^2)dxdy\\
    S_{xy}{shear}&=-\frac{1}{4\pi}\int_{x} \int_{y} (v_xB_{x}+v_yB_{y})B_z dxdy
\end{align}
Figure \ref{fig:PoyntingFip} (\ref{fig:PoyntingPip}) shows the time evolution of the two terms in the FI (PI) case. Figure \ref{fig:difference} shows the difference between the shear and the emergence term. Initially, in both cases (Figures \ref{fig:PoyntingFip} and \ref{fig:PoyntingPip}), the emergence term is larger than the shear term, which is reasonable, since the intense shearing starts after the bipolar region emerges at the photosphere. Eventually, the shear term becomes larger than the emergence term and it becomes the dominant term during the evolution of the system. After t= 80min, the oscillatory behaviour is due to eruptions. The formation and evolution of the eruptive structures will be discussed in a follow up paper. The difference between the two terms is larger in the PI case (\ref{fig:difference}). However, in the FI case with the same initial value for B, the difference of the two terms is even larger. The latter indicates that the inclusion of the partial ionization alone does not determine solely the amount of energy and/or flux, which is transferred at the photosphere and above an emerging bipolar region.

\begin{figure}[]
\centering
\includegraphics[width=\columnwidth]{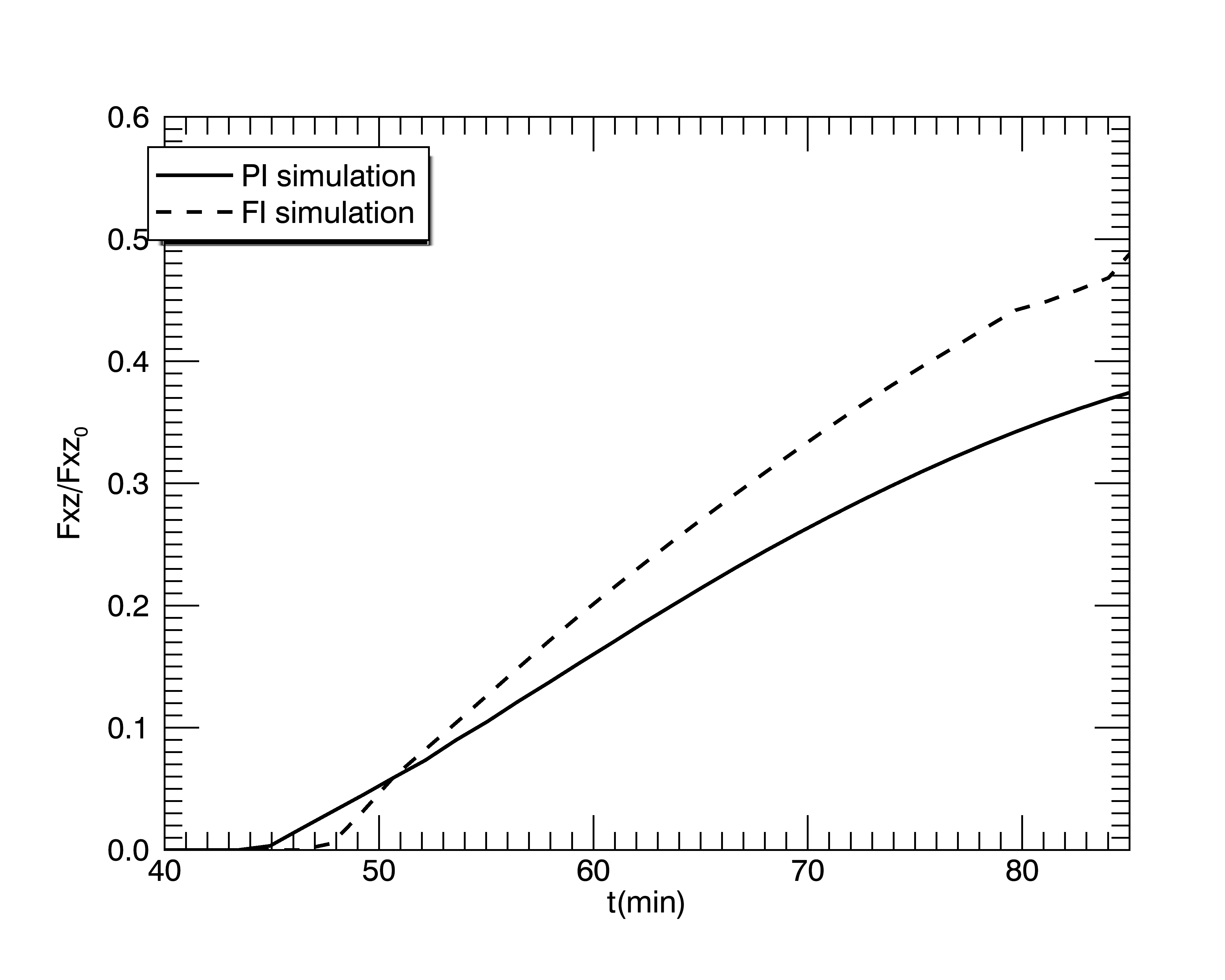}
\caption{Emerged normalized in plane flux above z=1.2Mm.}
\label{Fig:inplaneflux}
\end{figure}

We have also computed the normalized in plane flux as follows: 
\begin{equation}
    \frac{F_{xz}}{F_{xz_0}}=\frac{\int_{z>1.2Mm} \int_{x} B_{y} dx dz}{\int_{z} \int_{x} B_{y_0} dx  dz}
\end{equation}
where, $By_0$ is the axial component of the magnetic field inside the flux tube at t=0. The result is shown in figure \ref{Fig:inplaneflux}. Initially, for $t<=53$min, the in plane flux is larger for the PI case because the tube emerges and expands above the photosphere earlier. Eventually, it is the FI case where the in plane flux becomes bigger during the time evolution of the system. Therefore, it seems that a larger percentage of axial field (By) is transferred above the photospher in the FI case. 
\begin{figure}[]
\includegraphics[width=\columnwidth]{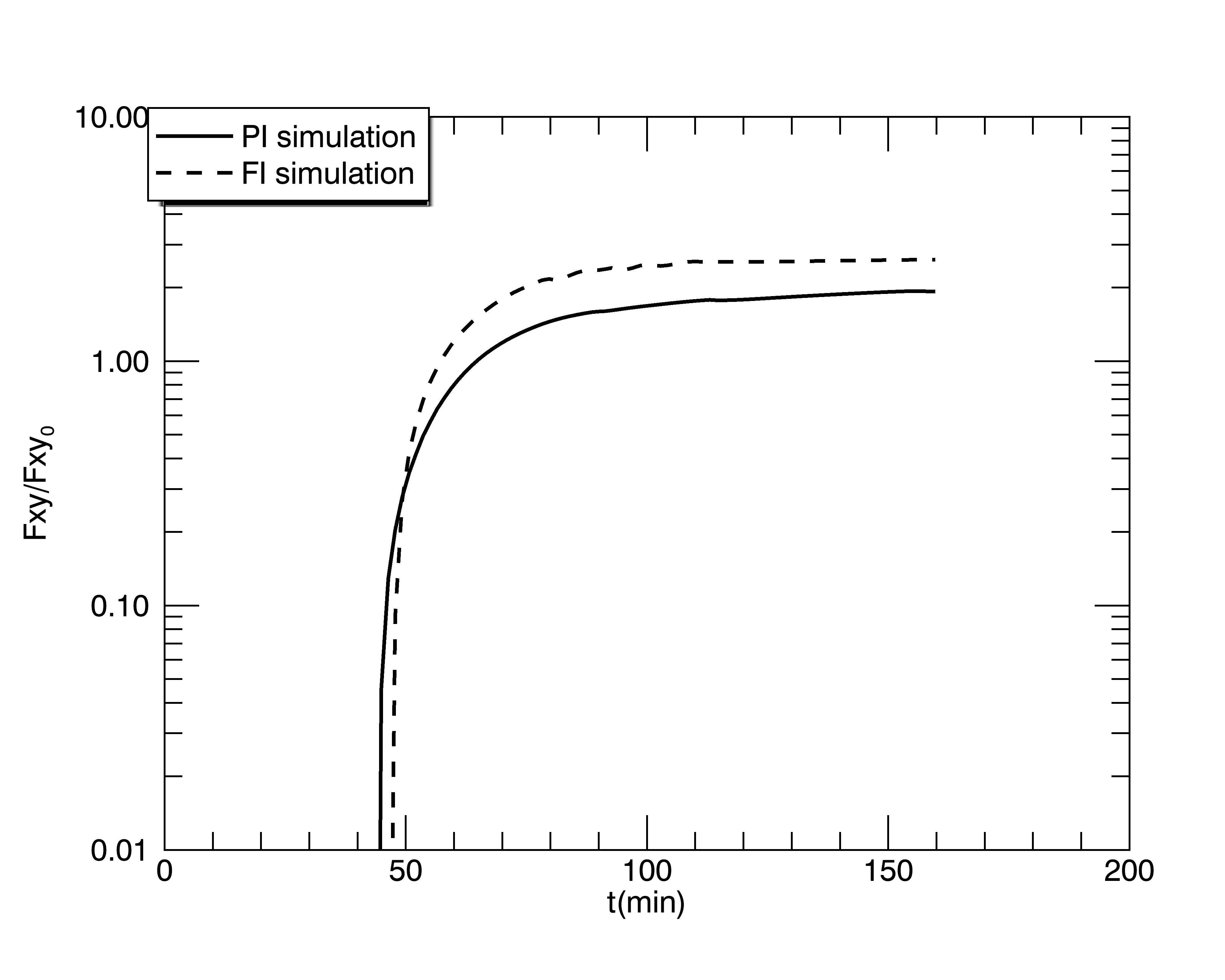}
\caption{Emerged vertical in plane flux at z=1.2Mm.}
\label{Fig:inplanevertflux}
\end{figure}
\begin{figure}[]
\centering
\includegraphics[width=\columnwidth]{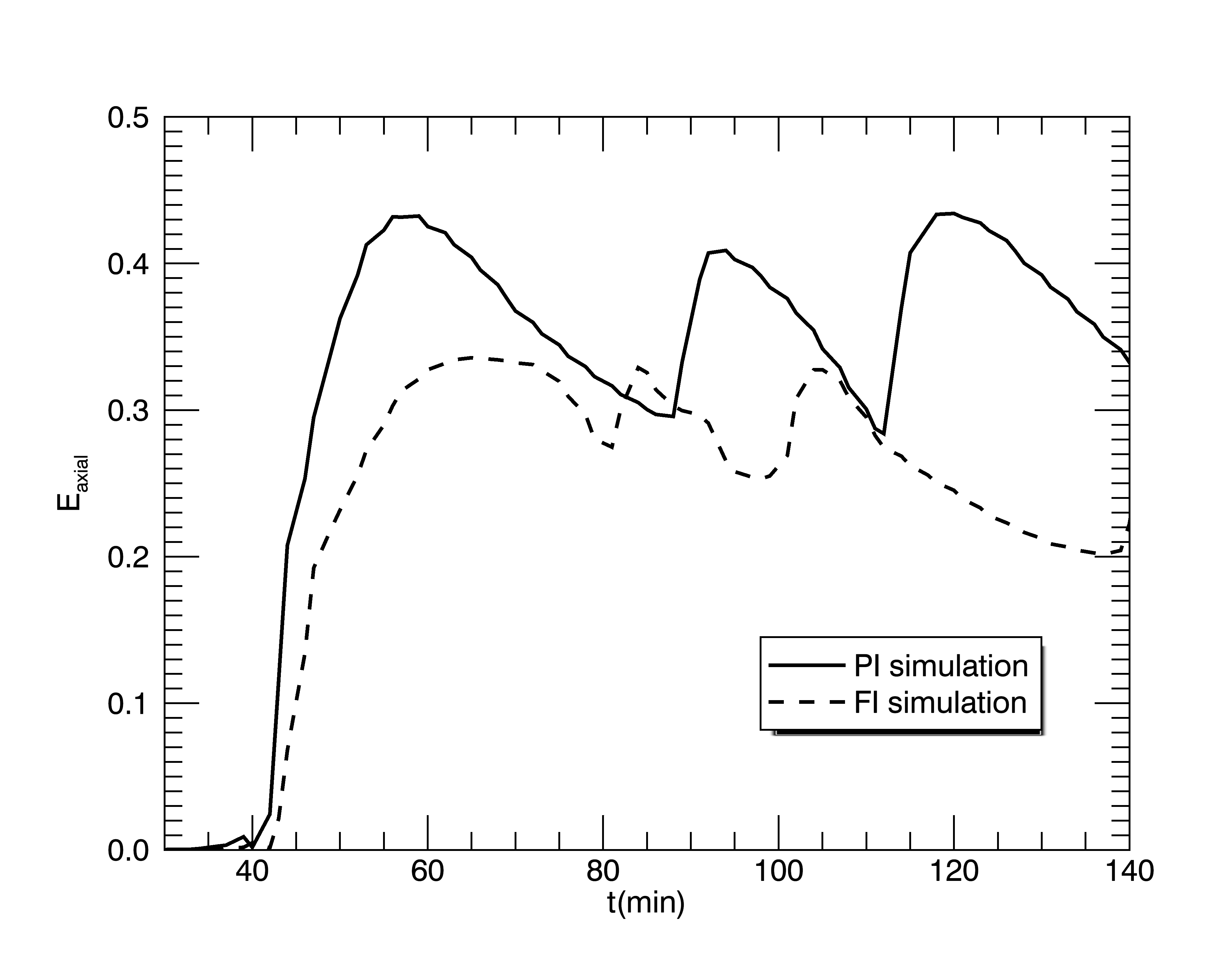}
\caption{Time evolution of $E_{axial}$ from $z>0.36 Mm$.}
\label{Fig:shear}
\end{figure}
We have also computed the unsigned normalized in plane vertical flux by using the following formula:
\begin{equation}
    \frac{F_{xy}}{F_{xy_0}}=\frac{\int_{x} \int_{y} |B_{z}| dx dy}{\int_{x} \int_{y} |B_{z_0}| dx  dy}
\end{equation}
We normalized it over the total initial in-plane vertical flux,  we used the whole domain for the x and y-axis, while z is equal to 1.2 Mm above the photosphere, this is on figure \ref{Fig:inplanevertflux}. In both cases, there is an initial rapid increase in the flux, which is due to the initial emergence of the magnetic flux tube at this height. This is followed by a slow increase and then saturation of the vertical flux, due to the fact that the available flux for emerging at this height has reached its limit. Overall, we find that the maximum value of the vertical flux above photosphere is similar in both cases (a bit larger in the FI case) and certainly the inclusion of PI does not lead to the emergence of more vertical flux at the solar atmosphere.

We have also computed an estimate of how much axial field energy(i.e due to the axial field component By) from the total magnetic energy, is transferred to the solar atmosphere. To do this, we calculate:
\begin{equation}
    E_{axial}=\frac{\int_{x} \int_{z>0.36} B_{y}^2 dx dz}{\int_{x} \int_{z>0.36} |B|^2 dx  dz}
\end{equation}
Figure \ref{Fig:shear} shows the time evolution of $E_{axial}$. We find that the values of $E_{axial}$ are higher (overall) in the PI case. This is mainly due to the initial conditions, where the subphotospheric field in both cases have the same plasma beta, but higher values for B and By in the PI case. We have performed one more simulation where the FI case has the same values for B and By with the PI case. In this third simulation, we find (not shown in this plot) that the $E_{axial}$ starts to increase earlier and it reaches higher values, at least at the first local maximum of its time evolution. In all  cases, the first peak corresponds to the initial emergence and expansion into the solar atmosphere. The oscillations after the first peak correspond to the eruptions of newly formed flux ropes due to the shearing along the PIL of the emerging bipolar regions.

\begin{figure*}[]
\centering
\includegraphics[width=1.085\textwidth]{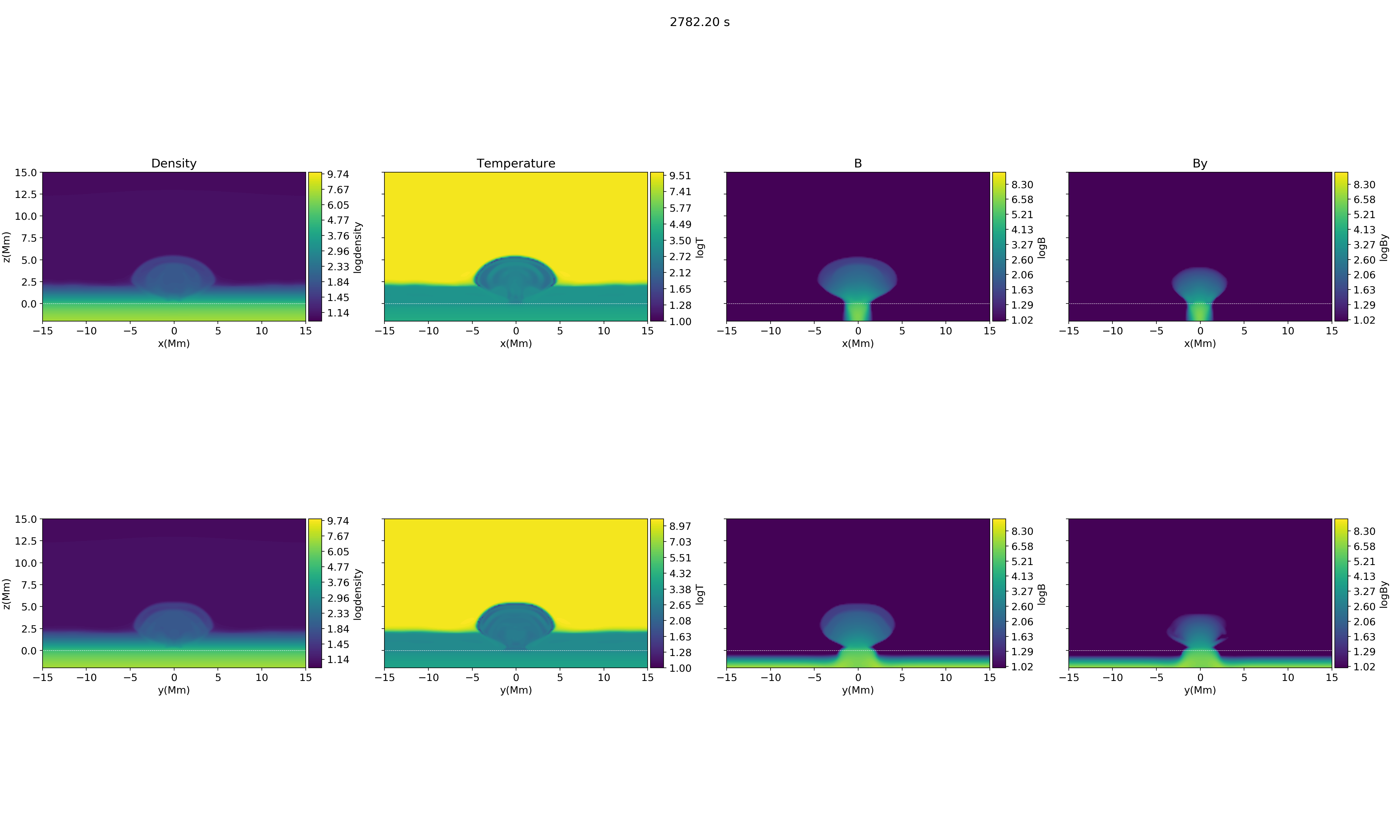}
\includegraphics[width=1.085\textwidth]{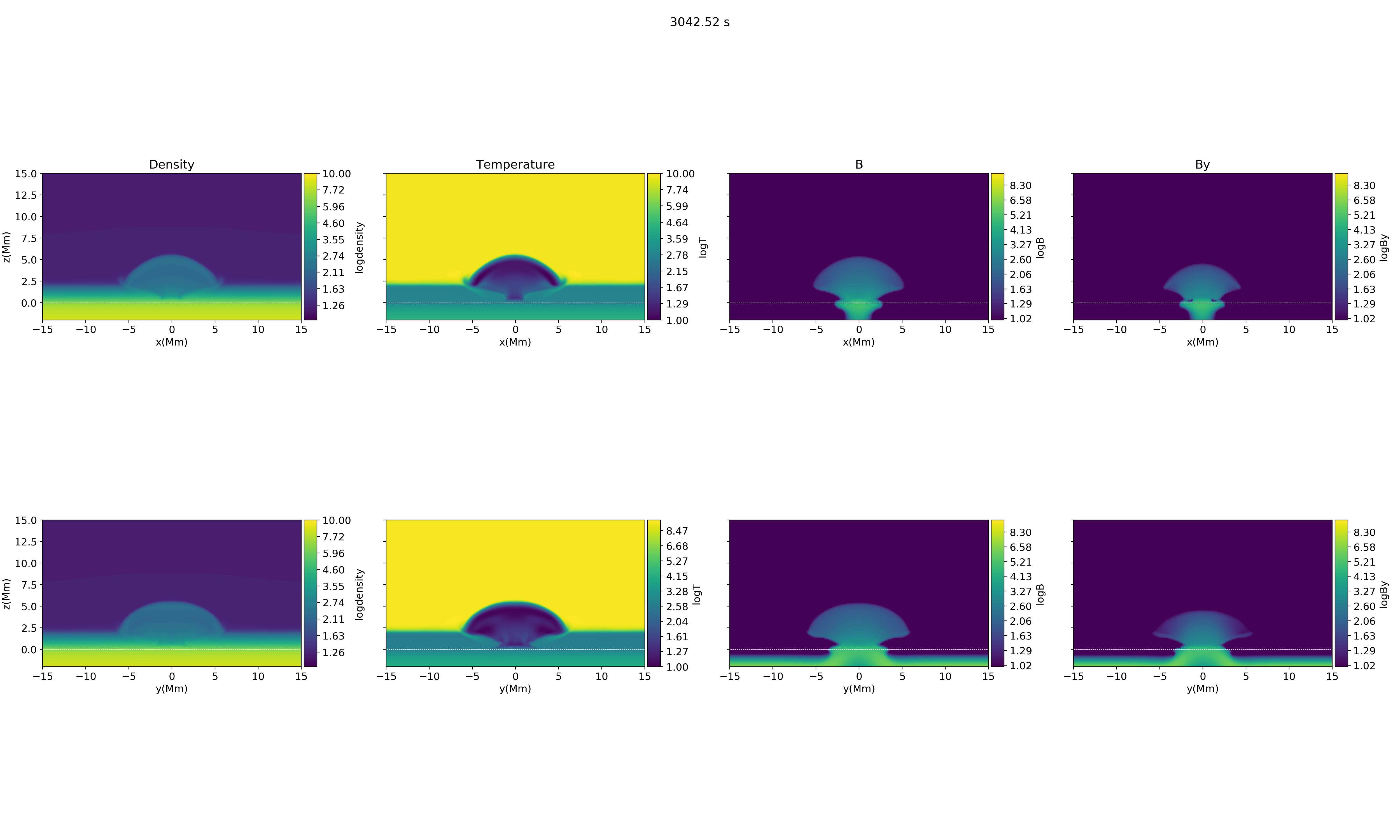}
\caption{Visualization of various quantities (temperature, density, total magnetic field strength (B) and axial magnetic field (By)), for the PI case (first two rows) and the FI case (last two rows), at the vertical xz-midplane (rows 1 and 3) and at the vertical yz-midplane (rows 2 and 4), at two different times. }
\label{Fig:dens}
\end{figure*}

\subsection{Magnetic flux emergence at solar atmosphere}

\begin{figure}
\includegraphics[width=\columnwidth]{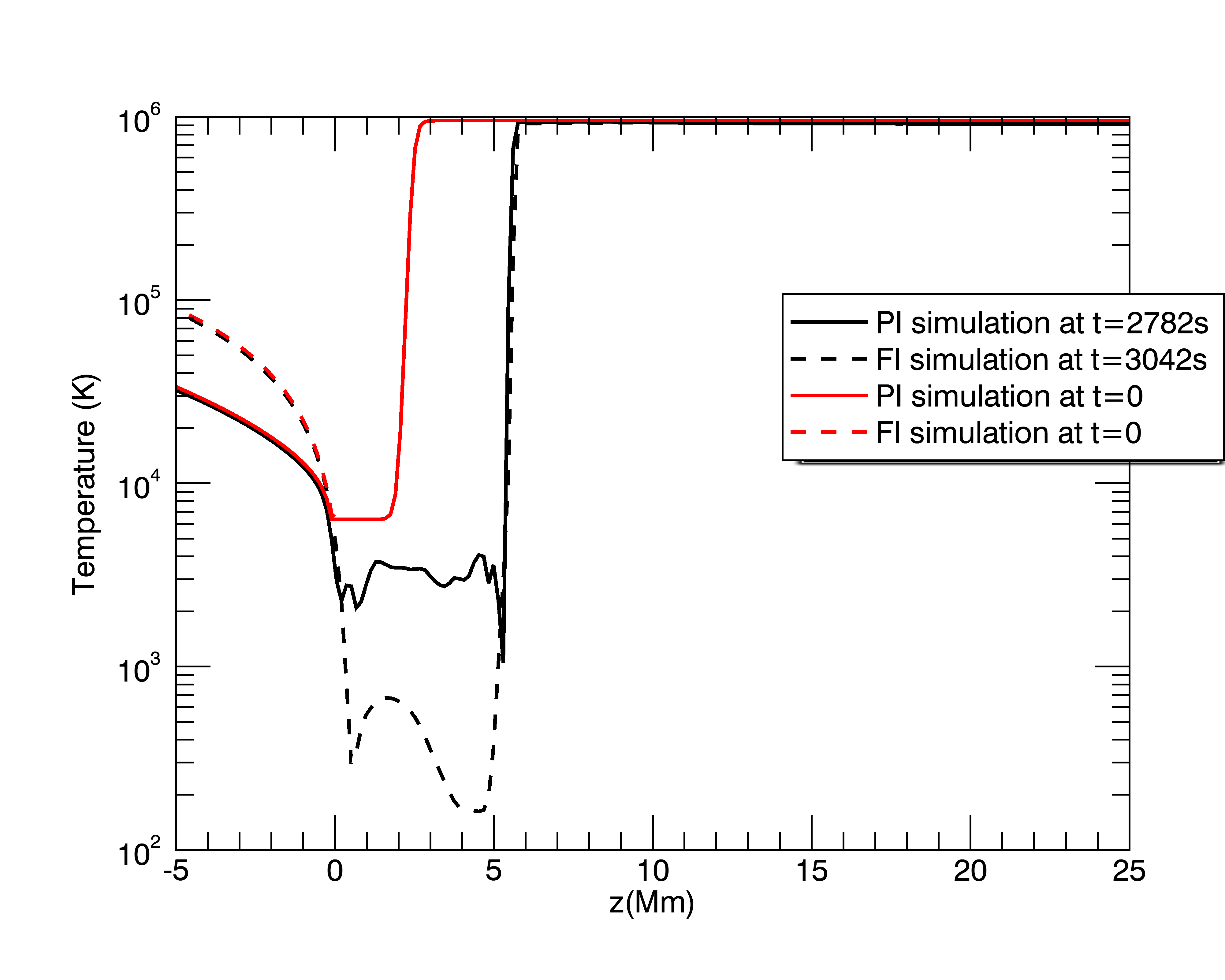}
\caption{The comparison of the temperature profiles on both simulations in the x=y=0 plane in t=2782s on FI and t=3042s on PI.}
\label{Fig:comparison}
\end{figure}

\begin{figure}
\includegraphics[width=0.485\textwidth]{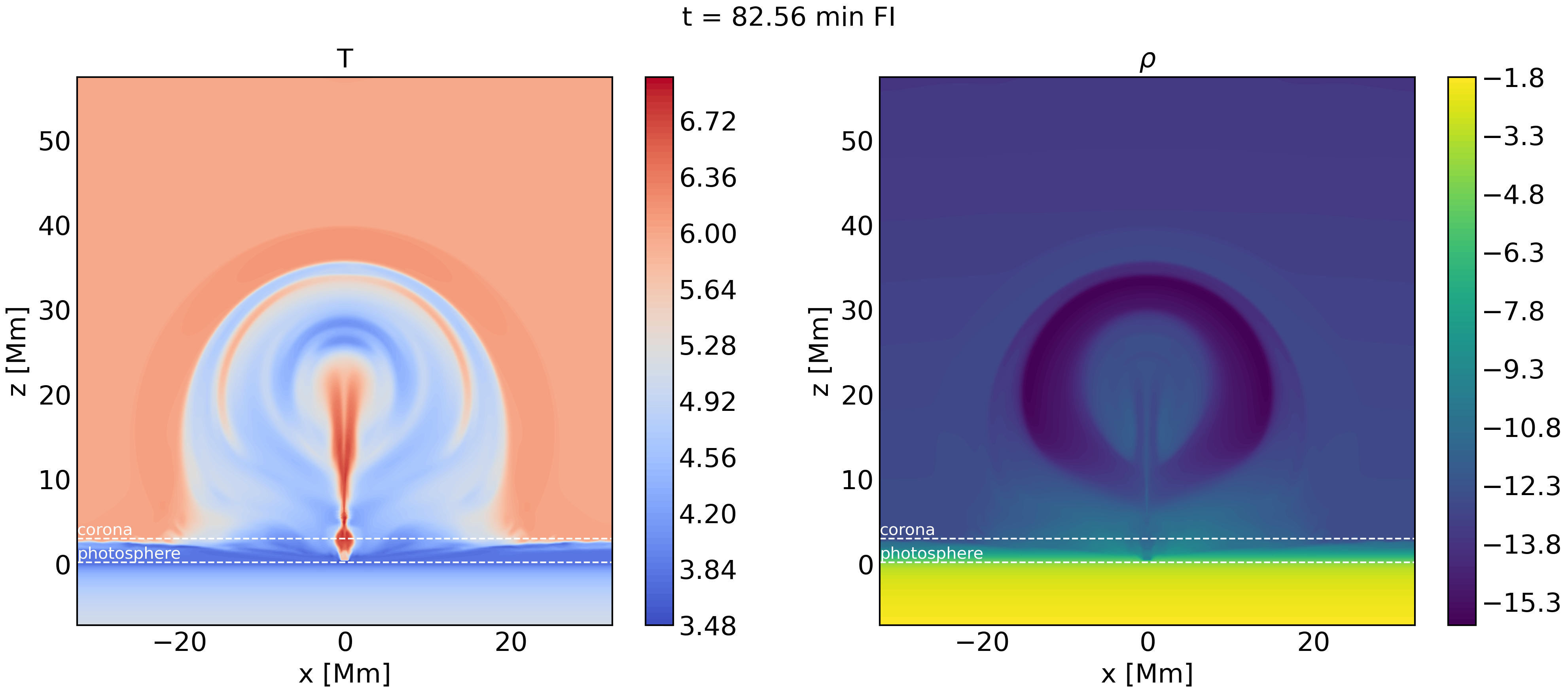}
\includegraphics[width=0.485\textwidth]{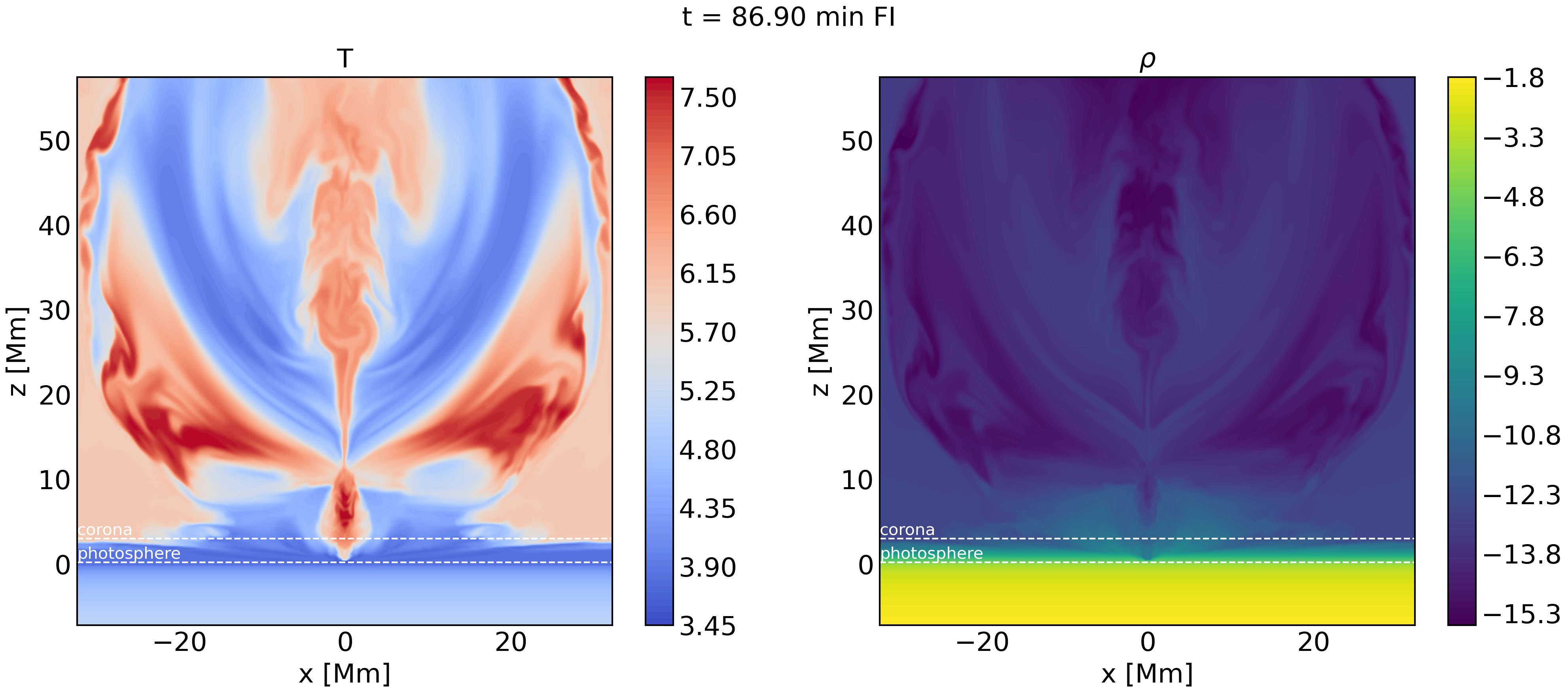}
\caption{Visualization of the logarithm of the temperature and the logarithm of density in the xz midplane for FI simulation in t=82.56 minutes and t=86.90 minutes. }
\label{Fig:eruption}
\end{figure}

\par Figure \ref{Fig:dens} shows the emergence of the magnetic flux into the solar atmosphere, in the two experiments (FI and PI cases) with the same initial plasma beta for the sub-photospheric magnetic field. We have chosen two different times (t=3042s for the FI case and t=2782s for the PI case), when the apex of expanding volume has reached the same atmospheric height.

By looking at the magnetic field (B and By), we find that in the FI case, the emerging field experiences a more profound horizontal expansion. This is due to two main reasons: a) because the emerging field intersects the photosphere in a more vertical manner and b) the apex of the emerging field carries more dense plasma into the atmosphere in the FI case. Both reasons are side effects of partial ionization as we have discussed earlier.

The visualization of the temperature (collumn 2) confirms two results, which have been reported in previous numerical simulations \cite{Leake_etal2006}. Firstly, the temperature inside the expanding field is lower than the background temperature due to the adiabatic expansion in the FI case. Secondly, in the PI case, the inclusion of the partial ionization reduces the adiabatic cooling. We have to highlight that this process does not heat the local plasma above nominal values. Therefore, perpendicular resistivity should be considered as a mechanism that reduces the heavy cooling from the adiabatic expansion, however it should not be considered (at least in this case) as a generic heating mechanism of the atmospheric plasma.

Figure \ref{Fig:comparison} displays a quantitative measurement of the adiabatic cooling reduction due to the effect of partial ionization. We plot the temperature along height, at the center of the computational domain (x=y=0) for both cases, when the emerging field has expanded into corona. The temperature of the plasma inside the expanding volume increases to about 3000 K (solid black line), but it is still lower than the nominal background value ( red solid/dashed lines).
\par In a similar manner to previous flux emergence simulations, we find that the emergence above the photosphere is followed by the formation of an erupting flux rope, due to the shearing and reconnection of magnetic fieldlines along the polarity inversion line. The eruption could be confined or ejective, depending on various parameters. Numerical
experiments without ambient magnetic field in the solar atmosphere, \citep[e.g.,][]{Fan_2001, Archontis_Torok2008, Archontis_etal2012},
have shown that the eruptions can be confined when the downward tension of the outermost fieldlines of the emerging field can keep the flux rope within the expanding volume of the emerging field. This occurs when, e.g., the flux rope does not have enough
free energy to push the outermost fieldlines strong enough and erupt in an ejective way. This could also occur when the numerical domain is small enough and the
boundary conditions are closed, so that the expansion of the emerging field and the overall evolution of the magnetic system is limited. Some of these experiments have shown that the inclusion of a pre-existing magnetic field in the solar atmosphere, which can reconnect effectively
with the emerging field, can lead to ejective eruptions of the flux ropes. Still, even
with a pre-existing magnetic field, if the free energy of the flux rope is relatively
small, the eruption is confined \cite{Leake2022ApJ...934...10L}. On the other hand, ejective eruptions
can occur without reconnection between the emerging and the pre-existing magnetic field,  \citep[e.g.,][]{Fan_2009, Archontis_etal2014, Syntelis_etal2017}. For instance, \cite{Syntelis_etal2017} showed that
successive eruptions can occur due to the combination of the torus instability and a
tether-cutting reconnection of the fieldlines enclosing the flux rope. Our simulations
have exactly the same initial conditions as the experiments by \cite{Syntelis_etal2017} and,
thus, the eruption mechanism (at least for the FI case) is the same.
Figure \ref{Fig:eruption} shows the first eruption of the FI case at $t=82.56$ minutes (first row) and $t=86.90$ minutes (second row). At $t=82.56$ min, the core of the flux rope is located around $x=0$ Mm and $z=25-30$ Mm. The plasma around the center of the flux rope is cool (top-left panel) and
dense (top-right panel). Underneath the flux rope, there is a hot vertical column of plasma. This column consists of a bidirectional pair of reconnection jets, which originate at the low corona (around $z=3$ Mm) and are formed due to tether-cutting reconnection of the fieldlines
surrounding the flux rope. At this stage of the evolution, the eruption is already
undergoing an ejective phase. At $t=86.90$ min (bottom row), the eruptive flux rope has left the upper part of the numerical domain, leaving behind a growing post-flare loop (its peak is located around $x=0$ Mm, $z=10$ Mm) and an intricate combination of hot and cool plasma in the corona. The onset and evolution of the eruptions for the FI and PI cases, and the similarities and differences between the two cases, will be discussed in detail ina forthcoming paper.

\section{Conclusions}
We have performed 3-D numerical simulations to investigate the effect of partial ionization on the magnetic flux emergence process. We have modified the single-fluid MHD equations to include the effects of neutral hydrogen on our simulations. This modification induced a different induction equation which included the collisions between the three different species and how these collisions affect the evolution of the magnetic field. 
\par We find that partial ionization affects the emergence of magnetic flux at the photosphere. The slippage effect is an imminent result of the ion-neutral collisions leading  to the dynamic movement of the ions through the plasma and  simultaneously  the couple of neutrals to the magnetic field. This slippage explains the slow horizontal expansion and the spindle-like shape of the emerging structure of the magnetic field.
\par Another result of the aforementioned phenomenon is the failed emergence of the FT axis in the photosphere. More precisely, the apex of the FT experiences the slippage effect which takes place if the ion-neutral collisions are sufficient to decouple the ions from the field. The axis on the other hand still “experiences” ion-neutral collisions but they are not sufficient in order to decouple the ions so the collisions there act like a deceleration device leading to the failed emergence of the axis to the solar surface. \par The ion-neutral collisions are also responsible for the less plasma that the PI axis is transferring to the solar atmosphere. This specific characteristic is the reason that the PI's apex starts to ascend towards the solar corona earlier than the FI case, although the emerging tubes in the two cases have the same plasma beta.\par Since our numerical simulations are fully three-dimensional, we have studied also the appearance of the emerging field around the PIL at the photosphere. We have found that in the PI case, the polarities adopt a more circular-like shape. This is because the dominant magnetic field component in this case is the vertical component, Bz. On the other hand, the magnetic tails are more apparent in the FI case, where the azimuthal component of the magnetic field is very strong.\par Previous studies \citep[e.g.][]{2012ApJ...747...87K,Martinez_Sykora_etal2015,Leake_etal2006,Leake_etal2013b} have indicated that partial ionization due to the induced perpendicular resistivity could be an important heating mechanism in the solar atmosphere. However, our simulation show that the inclusion of PI could reduce the intensive cooling  from the adiabatic expansion, yet it does not heat the plasma above the nominal background values.
\par We have also found that the amount of the emerged flux into the solar atmosphere is a little bigger in the PI simulation. However, this is mainly due to the fact that the initial sub-photospheric magnetic fields have the same plasma beta in both cases, but the initial field strength is bigger in the PI simulation. Therefore, the inclusion of PI can not determine (alone) the amount of flux transfer at and above the solar surface. Finally, our numerical experiments show that PI does not halt the formation of unstable structures at the PIL of the emerging region, which can eventually erupt into the outer solar atmosphere. To some extent, this result contradicts the assumption, based on the outcome of previous 2.5D studies \citep[e.g.,][]{Leake_etal2013b}, that PI may not help the creation of unstable coronal structures. The effect of PI on the recurrent eruptions and the comparison with the FI simulations, will be presented in a forthcoming study. 

\section*{Acknowledgements} 
The authors acknowledge support from the Royal Society
grant RGF/EA/180232. This work was also supported from the ERC synergy grant “The Whole Sun”. The authors express appreciation to Dr. D. Nóbrega-Siverio and Professor F. Moreno-Insertis. Their contributions,  during the ERC Synergy Grant workshop, "The Whole Sun", held at Paris-Saclay, provided helpful insights for our research. The work was supported by the High Performance Computing facilities of the University of St. Andrews “Kennedy”.

\bibliographystyle{apj}
\bibliography{bibliography}

\clearpage

\end{document}